\documentclass[fleqn,usenatbib]{mnras} 
\usepackage{newtxtext,newtxmath}
\usepackage[T1]{fontenc}
\DeclareRobustCommand{\VAN}[3]{#2}
\let\VANthebibliography\thebibliography
\def\thebibliography{\DeclareRobustCommand{\VAN}[3]{##3}\VANthebibliography}

\usepackage{amsmath}	
\usepackage{graphicx} 

\newcommand{\msun}{\mathrm{M_{\odot}}}
\newcommand{\rsun}{\mathrm{R_{\odot}}}
\newcommand{\st}{\textsc{StarTrack}}

\newcommand{\vej}{\ensuremath{v_{\rm ej}}}
\newcommand{\mbh}{\ensuremath{M_{\rm BH}}}
\newcommand{\mcl}{\ensuremath{M_{\rm cl}}}
\newcommand{\mcltot}{\ensuremath{M_{\rm cl,tot}}}

\def\porb{P_{\rm orb}}
\def\mcomp{M_{\rm comp}}
\def\rg{r_{\rm g}}

\title[The Gaia~BH-like binaries]{The enigmatic origin of two dormant BH binaries: Gaia BH1 and Gaia BH2.}

\author[I. Kotko et al.]{
Kotko,I.,$^{1}$\thanks{E-mail: ikotko@camk.edu.pl}
Banerjee,S.,$^{2,3}$
Belczynski,K.,$^{1}$\thanks{Deceased}
\\
$^{1}$Nicolaus Copernicus Astronomical Centre
of the Polish Academy of Sciences, ul. Bartycka 18, 00-716 Warszawa, Poland \\
$^{2}$Helmholtz-Instituts für Strahlen- und Kernphysik (HISKP),\\ Nussallee 14-16, D-53115 Bonn, Germany \\
$^{3}$Argelander-Institut für Astronomie (AIfA), Auf dem Hügel 71, D-53121, Bonn, Germany \\
}
\date{}

\begin{document}

\label{firstpage}
\pagerange{\pageref{firstpage}--\pageref{lastpage}}
\maketitle

\begin{abstract}
The two systems, namely, Gaia~BH1 and Gaia~BH2, that have been confirmed as dormant (i.e., no X-ray emission detected) black hole (BH) - low-mass star binaries in the latest Gaia mission data release (DR3) are intriguing in the context of their formation and evolution. Both systems consist of $\sim9\,\msun$ BH and $\sim1\,\msun$ star orbiting each other on a wide, eccentric orbit ($e\sim 0.5$). We argue that formation of such Gaia~BH-like systems through the isolated binary evolution (IBE) channel, under the standard common envelope assumptions, and from dynamical interactions in young massive and open clusters are equally probable, and that the formation rate of such binaries is of the order of $10^{-7}\,\msun^{-1}$ for both channels. We estimate that, according to our models, there are at most $\sim900$ detectable Gaia~BH-like binaries in the Milky Way thin disc.
What plays an important role in formation of Gaia~BH-like systems via the IBE channel is the mutual position of the natal kick velocity vector and the binary angular momentum vector. We find that natal kicks with a median magnitude of $\sim40$ km/s are preferred for the formation of Gaia~BH1-like binaries. Approximately $94\%$ of those binaries are formed with the BH spin misaligned to the orbital axis by less than $40^{\circ}$. Gaia~BH2-like binaries form if the low velocity natal kick (of median magnitude $\sim20$ km/s) is directed within $15^{\circ}$ about the orbital plane.
In addition to natal kick, we also discuss the influence of tidal interaction and the adopted common envelope $\lambda_\mathrm{ce}$~parameter prescription on the evolution of Gaia~BH-like binaries.
We follow the subsequent evolution of the binaries, once formed as Gaia~BH1 and Gaia~BH2 systems, to investigate their connection with the low-mass X-ray binary population.
\end{abstract}




\section{Introduction}

The dormant black hole (BH) binaries are difficult to observe as they do not emit X-ray radiation and can be discovered only by astrometric and/or spectroscopic analysis. Although such systems were predicted to exist already in the 1960s \citep{Guseinov66}, the new generation of the astronomical instruments and new data analysis methods have allowed to identify the first candidates for such objects only recently. To our knowledge, the first strong candidate for a binary system with a heavy companion unseen in electromagnetic spectrum was found in the globular cluster NGC 3201 \citet{Giesers18}.  Within the last two years, another four systems have been discovered: 
{\em(i)} binary HD 130298 where the minimum mass of an unseen companion has been estimated to be $7.7\pm1.5$ \(\msun\) \citep{Mahy22},
{\em(ii)} VFTS 243 found in Large Magellanic Cloud which is a binary of a massive star and most probably a black hole companion with mass $M=10.1\pm2.0$ \(\msun\) \citep{Shenar22}, and 
{\em(iii)} two systems which were identified in the most recent Gaia catalogue DR3 for which there is the strong evidence that they host a non-accreting BH and a low-mass companion star \citep{Chakrabarti2023,El-Badry23_GaiaBH1,El-Badry23GaiaBH2,Tanikawa23}. 
In our studies, we focus only on the last two systems and, following \citet{El-Badry23GaiaBH2}, we refer to those binaries as Gaia~BH1 and Gaia~BH2.\\
Once Gaia~BH1 and Gaia~BH2 had been detected, questions and speculations arose about the origin of those systems. Several studies have already been made which point towards dynamical interactions in young stellar clusters as the most probable scenario to create a BH-low-mass star system having a moderately wide and eccentric orbit \citep{El-Badry23_GaiaBH1,Tanikawa2023,Rastello23_gaiaBH1_form,DiCarlo23_GaiaBH_form}. However, as we show in this work, the isolated binary evolution (IBE) channel may contribute to the total population of dormant BH binaries at a rate that is comparable to that for the dynamical-interaction channel. The important factors that impact the number of the binaries of interest born through the IBE channel is the magnitude and the geometry of the natal kick imparted to the newborn compact object after the supernova (SN) explosion. One can infer about the natal kick from the peculiar velocity\footnote{The peculiar velocity corresponds to systemic velocity in the model, as is explained in Sec.~\ref{sec:IBE_Method}.} and the orbital parameters of the binary system. \\
There are two components that change the systemic velocity of a binary after an SN event in the model:
{\em(i)} the Blaauw kick that impacts the velocity of the binary centre of gravity due to symmetric mass ejection during the SN \citep{Blaauw61},
{\em(ii)} the natal kick that is received by the compact object due to asymmetries of mass ejecta and/or neutrino emission during the SN \citep[e.g.]{Chugai84,Janka13,Janka17}. 
The observations of proper motions of pulsars and X-ray binaries give constrains on the natal kicks received by neutron stars (NSs) or BHs upon their formation. The Maxwellian distribution with dispersion velocity $\sigma=265$ km/s, as derived by \citet{Hobbs05} from observations of Galactic pulsars, is widely used in the binary population synthesis simulations. This distribution is applied also to BHs but with additional assumption that the kick velocity is reduced by the amount of matter that falls back, being pulled by the gravitational field of the BH forming during the SN. However, recent observations indicate that this approach may underestimate the number of lower velocity systems and overestimate the number of the systems in higher velocity range and that it should be revisited \citep{Atri19,O'Doherty23,Zhao2023}. The magnitude of the natal kicks is only one part of the puzzle, the other is the kick velocity vector orientation. In the codes based on BSE \citep{Hurley00}, the direction of a natal kick is drawn from an isotropic spherical distribution. The question is whether the specific direction of the natal kick, that a BH receives in an SN explosion, could lead to the formation of Gaia~BH1 and Gaia~BH2 (under the standard assumptions of isolated binary evolution model). The reason we may attempt to answer this question for the two systems of interest is that Gaia~BH1 and Gaia~BH2 are not impacted by the orbit changes due to the mass transfer events after the SN. Gaia~BH1 has the orbital separation and eccentricity as it was shaped by SN and Gaia~BH2 orbit is only modulated by weak tidal forces.\\
Our work consists of two main parts: in Sec.~\ref{sec:IBE}, we consider the IBE channel and follow the evolution of the binaries which we classify as Gaia~BH1-like and Gaia~BH2-like and find the probability of formation of such binaries from the IBE. We then run the evolution of two well-matching Gaia~BH1 and Gaia~BH2 binaries until the Hubble time and investigate what information Gaia~BH1 and Gaia~BH2 may bring us about the natal kicks. In Sec.~\ref{GaiaBH_clusters}, we present and discuss the formation rates of Gaia~BH-like binaries from young massive and open star clusters.
\begin{table}
    \centering
    \caption{The parameters of  Gaia~BH1 and Gaia~BH2: the orbital period $P_\mathrm{orb}$ in days, the eccentricity $e$, BH mass $M_\mathrm{BH}$ and companion star mass $M_{\star}$ in solar units, companion evolutionary stage and age in Gyrs, adopted from \citet{El-Badry23_GaiaBH1,Chakrabarti2023,El-Badry23GaiaBH2}. The peculiar velocity at birth $v_\mathrm{pec}$ is taken from Table~3 in \citet{Zhao2023}.} 
    \label{Table:Gaia param}
    \resizebox{\hsize}{!}
    {
	\begin{tabular}{c|c c} 
         \hline
                                     & Gaia~BH1     & Gaia~BH2\\
          \hline   
            $P_\mathrm{orb}$ [d]   & $185.52\pm0.08$  & $1276.7\pm0.6$ \\
            $e$                    & $0.439^{0.004}_{-0.003}$ & $0.5176\pm0.0009$ \\
		$M_\mathrm{BH}$ [$\msun$]  & $9.326^{0.216}_{-0.209}$    & $8.94\pm0.34$ \\
		$M_{\star}$ [$\msun$]      & $0.93\pm0.05$    & $1.07\pm0.19$  \\
            Star type                  & MS          & RG          \\
            Age [Gyr]            & $>4.0$     & $8.0-12.0$ \\  
        $v_\mathrm{pec}\,[\mathrm{km/s}]$    & $71.3^{+9.2}_{-10.8}$    & $34.1^{+5.0}_{-5.1}$ \\
            
            \hline
	\end{tabular}
    }
\end{table}
\section{Gaia BH binaries from isolated binary evolution}\label{sec:IBE}
\subsection{Method}\label{sec:IBE_Method}
We performed the population synthesis calculations using $\st$. The detailed description of the code can be found in \citet{Belczynski02,Belczynski08,Belczynski16}. Here we highlight the assumptions that are most important in the context of the present paper. \\
The adopted initial mass distribution is a power law function with three exponents \citet{Kroupa93,Kroupa01}: $\alpha_1=-1.3$ for stars of masses $0.08<M<0.5\,\msun$, $\alpha_2=-2.2$ for $0.5\leq M<1.0\,\msun$ and $\alpha_3=-2.3$ for stars more massive than $1.0\,\msun$. To maximise the effectiveness of calculations we focused on the systems that may be the progenitors of Gaia~BH binaries: the mass of BH-progenitor star was drawn from the mass range $18.0-200.0\,\msun$ and the mass of its companion from the range $0.2-1.5\,\msun$. The orbital period ($P/\mathrm{orb}$) (in units of days) and eccentricity $e$ were drawn from the distributions:  $f(\log{P/\mathrm{orb}})=(\log{P/\mathrm{orb}})^{-0.55}$  with $\log{P/\mathrm{orb}}$ in the range $[0.15,4.0]$ and $f(e)=e^{-0.42}$ with $e$ in the range $[0.0,0.9]$ respectively \citep{Sana12,Sana13}.\\
In our calculations the common envelope phase is parameterized by CE parameter $\alpha_\mathrm{CE}=\alpha_{\mathrm{ce}}\times\lambda_{\mathrm{ce}}$ where $\lambda{\mathrm{ce}}$ is a parameter describing the donor central concentration \citep{deKool1990_lambda} and $\alpha_{\mathrm{ce}}$ parameterizes the efficiency at which the orbital energy is transferred into the envelope \citep{Webbink84}. We assume the standard value $\alpha_{\mathrm{ce}}=1.0$ and for $\lambda_{\mathrm{ce}}$ we use the fits from \citet{Xu2010,Xu2010_Erratum} and \citet{Wang2016} (see Sec.\ref{ssec:discussion_lambda} for discussion).\\
The massive stars (of type O/B) stellar winds were described by the equations for radiation driven mass loss from \citep{Vink2001} which take into account the Luminous Blue Variable mass loss \citep{Belczynski10}. For Wolf-Rayet stars we took into account the wind clumping \citep{Hamann_Koesterke98} and the metallicity dependence of the winds \citep{Vink_deKoter05}. The winds of less massive stars were calculated according to the formulae of \citet{Hurley00}. \\We assumed the delayed supernova engine \citep{Fryer12} in our calculations. 
There are two components that make up the final kick received by the system in SN explosion: the kick from asymmetric neutrino and/or ejecta emission (natal kick) and the Blaauw kick resulting from the symmetric envelope ejection from the exploding star.
We tested two distributions of the natal kick velocities: i) the widely used Maxwellian distribution with $\sigma=265\,\mathrm{km\,s^{-1}}$  \citep{Hobbs05} which for BH is decreased by the fallback of the fraction of the stellar envelope that is initially ejected in SN explosion, this fraction is defined by the fallback parameter $f_{\mathrm{fb}}$ which takes the values between $0$ and $1$ (model V1), ii) the two-component Maxwellian with $\sigma_1=21.3\,\mathrm{km\,s^{-1}}$ for low-velocity component and $\sigma_2=106.7\,\mathrm{km\,s^{-1}}$ for high-velocity component derived from the sample of 85 BH and NS X-ray binaries by \citet{Zhao2023} also decreased by the fallback for BHs (model V2). The directions of the kicks are drawn from isotropic distribution. For each system the drawing of natal kick magnitude and orientation was repeated $200$ times.\\
The systemic velocity $v_\mathrm{sys}$ calculated in $\st$ is the velocity of the binary centre of mass resulting from Blaauw kick and natal kick. It is assumed that the center of mass is initially at rest. $v_\mathrm{sys}$ is consistent with the potential peculiar velocity at birth $v_\mathrm{p,z=0}$ \citep{Atri19}, that is the velocity that the binary gained from SN which caused its motion out of the Galactic plane. The potential peculiar velocity is inferred from the observational present-day peculiar velocity (for details see \citet{Atri19,Zhao2023}).\\
We position the orbital plane of a binary before the first SN on x-y plane. All three vectors, the orbital angular momentum and the spins of a star and BH, are aligned and pointing in $z$ direction. Setting binary in such position does not influence our results while it makes the analysis of the post-SN relative orientation of those vectors much easier. The position of the exploding star on the orbit is found by means of mean anomaly $M$, eccentric anomaly $E$, eccentricity $e$ and the time of periapsis passage $\tau$.\\
Taking into account the age constrains derived from the observations of the luminous component in Gaia~BH1 and Gaia~BH2 we assumed that both binaries are $\sim4$ Gyr old \citep{El-Badry23_GaiaBH1,El-Badry23GaiaBH2} and, therefore, were formed in the environment of metallicity $Z=0.7Z_{\odot}$ \citep[see Fig.3 in][]{Olejak20} where$Z_{\odot}=0.014$ \citep{Asplund09}.\\
It is hard to expect that one may get many systems of the parameters exactly corresponding to those inferred from observations of Gaia~BH1 and Gaia~BH2 in the simulations. Therefore, by putting constrains on the orbital separation $a$, the eccentricity $e$, the BH mass $M_\mathrm{BH}$, the companion star mass $M_2$ and the companion star evolutionary stage, we defined three groups of binaries which are of our interest: Gaia~BH1-like, Gaia~BH2-like and Gaia~BH-like. Gaia~BH1- and Gaia~BH2-like are the systems assumed to correspond to observed Gaia~BH1 and Gaia~BH2 while Gaia~BH-like is the general group of all systems which may be classified as dormant BH binaries. 
To investigate the interconnection between the natal kicks and the post-SN quantities: $a$, $e$, $v_\mathrm{sys}$, orbital angular momentum and BH spin vectors orientation we defined the fourth group - the "optimistic" Gaia BH-like binaries - for which we allow the wider range of $a$, $e$ and both components masses. The classification criteria for each group are given in Table~\ref{Table:Gaia-like param}. \\
We define two angles which help to complement our analysis: 
{\em (i)} $\gamma$ - a misalignment between the binary angular momentum vector and the natal kick velocity vector, and {\em (ii)} $\theta$ - a misalignment between the binary angular momentum vector and post-SN BH spin vector.\\

\begin{table}
    \centering
    \caption{The parameters of the binaries that we classify as Gaia~BH1-, Gaia~BH2-, and Gaia~BH-like in our standard models V1 and V2. We take into account both components masses ($M_\mathrm{BH}$ and $M_2$), the evolutionary stage of the companion (main sequence (MS) /red giant (RG)), the orbital separation $a$, the eccentricity $e$ and the systemic velocity $v_\mathrm{sys}$ (km/s).} 
    \label{Table:Gaia-like param}
    \resizebox{\hsize}{!}
    {
	\begin{tabular}{c|c c c c} 
         \hline
                                     &                  &               & standard & optimistic \\
                                     & Gaia~BH1-like     & Gaia~BH2-like  & Gaia BH-like & Gaia BH-like\\
          \hline                       
		$M_\mathrm{BH}$ [\(\msun\)] & 8.0-12.0    & 7.0-11.0 & <15.0 & < 15.0\\
		$M_2$ [\(\msun\)]           & 0.8-1.2     & 0.8-1.2  & 0.8-1.2 & < 1.5 \\
            Companion star           & MS          & RG     & MS/RG & MS/RG \\
            $a$ [\(\rsun\)]            & 200-400     & 950-1150 & 200-1200 & 100-1500\\  
            $e$                      & 0.4-0.6     & 0.4-0.6  & 0.3-0.7 & 0.1-0.9\\
    $v_\mathrm{sys}\,[\mathrm{km/s}]$    & no limit    & no limit  & no limit & no limit\\
            
            \hline
	\end{tabular}
    }
\end{table}

\subsection{Results}\label{sec:results}

\subsubsection{The formation rate of Gaia~BH-like systems from IBE}\label{ssec:results_formation_rates}

The formation rate of Gaia~BH-like binaries depends on the criteria according to which one classifies the binary as Gaia~BH-like and on the adopted natal kick velocity distribution and it may differ as much as three orders of magnitude as can be seen from Table~\ref{Table:IBE_rates}. The formation rate for model V1 is an order of magnitude smaller than formation rates for model V2. There is also one order of magnitude difference between the standard and optimistic groups in both models: for optimistic and standard Gaia~BH-like groups in model V2 the formation rate is $\sim1.4\times10^{-7}\,\msun^{-1}$ and $\sim1.0\times10^{-8}\,\msun^{-1}$ respectively and in model V1 it is $\sim1.6\times10^{-9}\,\msun^{-1}$ and $\sim1.1\times10^{-8}\,\msun^{-1}$ respectively.\\
The formation rates of  Gaia~BH1- and Gaia~BH2-like systems, which are the two binary categories with the most stringent classification criteria among the binary types defined in Table~\ref{Table:Gaia-like param}, are on order of $10^{-10}\,\msun^{-1}$. The formation rate of Gaia~BH2-like systems is almost an order of magnitude higher in model V2 than in model V1, while the formation rate of Gaia~BH1-like binaries does not depend on the natal kick distribution adopted (model V1 vs V2).\\
Additionally, to make the comparison with the results of dynamical channel described in Sec.~\ref{res_clusters}, we calculated the general population of BH-star binaries with the only constrain imposed on the companion star mass ($\mcomp<3.0\,\msun$) and we found that the formation rate of such binaries from IBE channel is $2.2\times10^{-6}\,\msun^{-1}$.

\begin{table}
    \centering
    \caption{The formation rates of Gaia~BH-like binaries for different classification criteria and natal kick velocity distributions V1 and V2. The criteria adopted to calculated the formation rate for case "V2 low-mass comp" correspond to the population of BH-star binaries with low-mass companion ($\mcomp<3\,\msun$) from young massive and open clusters described in Sec.~\ref{res_clusters}. We present the formation rates of Gaia~BH1- and Gaia~BH2-like binaries in model V1 and V2 in the last three rows.}
    \label{Table:IBE_rates}
    {
	\begin{tabular}{lc} 
         \hline
          Case             & Rate [$\msun^{-1}$] \\
          \hline  
          V1  standard &     $\sim1.6\times10^{-9}$ \\
          V1  optimistic &   $\sim1.1\times10^{-8}$ \\
	   V2  standard &     $\sim1.0\times10^{-8}$  \\
          V2  optimistic &   $\sim1.4\times10^{-7}$  \\
          V2 low-mass comp & $\sim2.2\times10^{-6}$  \\
          V1  standard &     $\sim1.6\times10^{-9}$ \\
          V1 $\&$ V2 Gaia~BH1 &$\sim3.2\times10^{-10}$ \\
          V1 Gaia~BH2 &     $\sim1.4\times10^{-10}$ \\
          V2 Gaia~BH2 &     $\sim9.6\times10^{-10}$ \\
            \hline
	\end{tabular}
    }
\end{table}

Among Gaia~BH1-like an Gaia~BH2-like binaries we present two which are our closets match to the observational systems in Table~\ref{Table:Gaia_examples_param}.

\begin{table}
    \centering
    \caption{The parameters of the two Gaia~BH1-like and Gaia~BH2-like binaries that we find closely matching the observed systems: the orbital period $P_\mathrm{orb}$ in days, the eccentricity $e$, BH mass $M_\mathrm{BH}$ and companion star mass $M_{\star}$ in solar units and the systemic velocity $v_\mathrm{sys}$.} 
    \label{Table:Gaia_examples_param}
    \resizebox{\hsize}{!}
    {
	\begin{tabular}{c|c c} 
         \hline
                                     & Gaia~BH1-like    & Gaia~BH2-like\\
          \hline   
            $P_\mathrm{orb}$ [d]   & $181.07$  & $1591.63$ \\
            $e$                    & $0.48$ & $0.56$ \\
		$M_\mathrm{BH}$ [$\msun$]  & $8.22$   & $8.55$ \\
		$M_{\star}$ [$\msun$]      & $1.38$    & $1.14$  \\
        $v_\mathrm{pec}\,[\mathrm{km/s}]$    & $57.73$ & $35.57$ \\           
            \hline
	\end{tabular}
    }
\end{table}

\subsubsection{The evolution of Gaia~BH1- and Gaia~BH2-like systems}\label{ssec:results_evolution}

\begin{figure}
\centering
\resizebox{\hsize}{!}
{\includegraphics[width=\textwidth,angle=270]{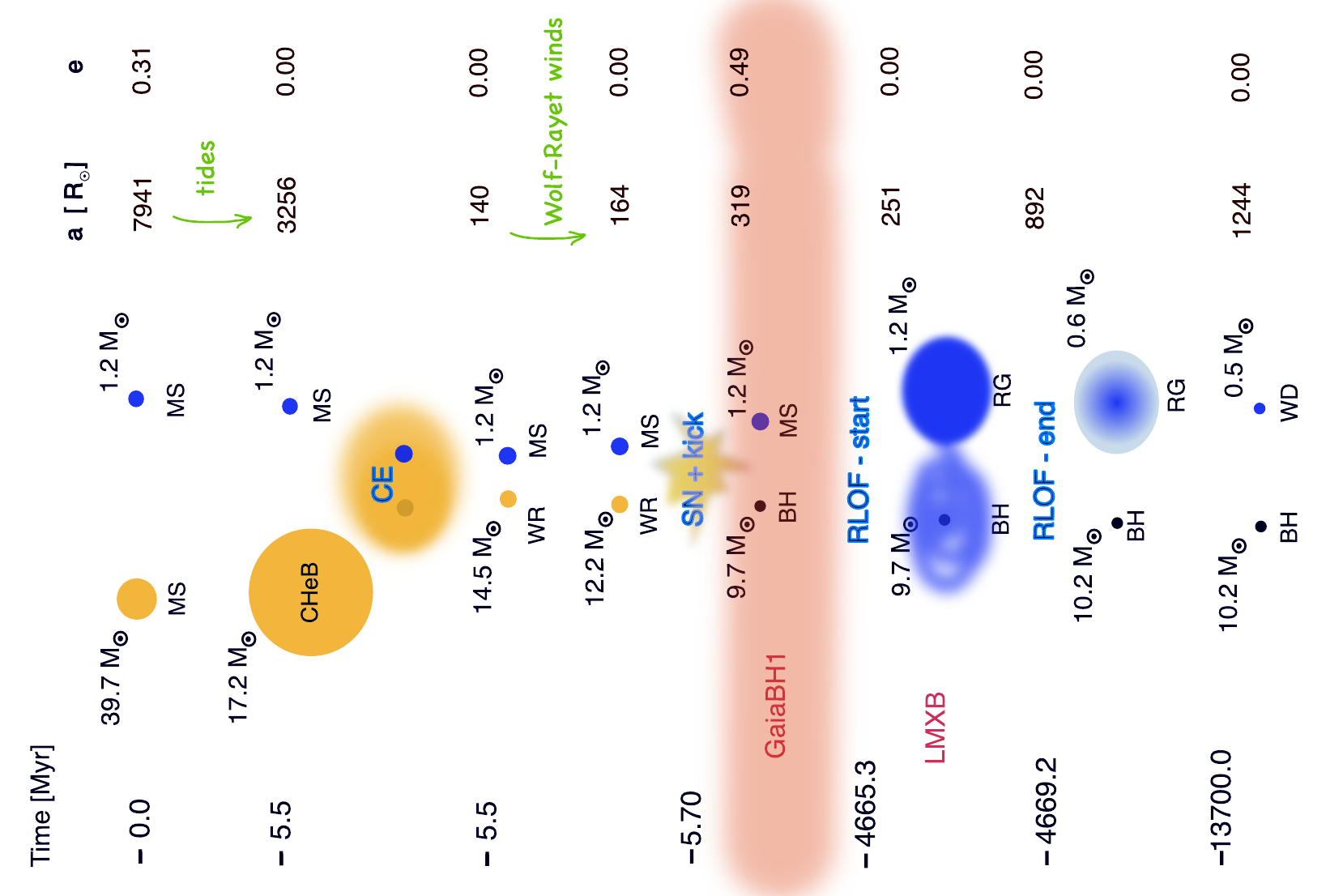}}
\caption{The evolution of Gaia~BH1-like binary from the isolated binary evolution channel. The chosen example has the parameters close to the observational ones.}
\label{Fig:GaiaBH1_evol}
\end{figure}

The conditions that we set on BH masses for Gaia~BH1- and Gaia~BH2-like binaries limit the ZAMS masses of BH progenitors to the range $M_\mathrm{ZAMS}\sim34.0-40.0\,\msun$ (for $Z=0.01$). We find that the initial orbital separations may reach up to $\sim36\,000\,\rsun$ for very high initial eccentricities ($e\sim0.9$). For the moderate initial eccentricities $e<0.4$ the initial orbital separations are in the range $\sim6000-9000\,\mathrm{R}_{\odot}$.\\
Whatever their starting parameters are, all Gaia~BH1-like binaries follow the same evolutionary path which we describe on one example also shown on Fig.~\ref{Fig:GaiaBH1_evol}. The evolution starts with two main sequence stars with masses $39.7\,\msun$ and $1.2\,\msun$ on a wide ($a\sim7\,941\,\rsun$), moderately eccentric ($e=0.31$) orbit. In the course of the nuclear evolution of the more massive star (the primary) the orbit widens due to the strong stellar winds and it reaches $a\sim8\,542\,\rsun$ at the time when the star leaves the Hertzsprung gap and starts to burn helium in its core. The still expanding radius of the core helium burning star (CHeB) is large enough ($\sim1550\,\rsun$) that the tides start to act between the two stars causing the orbit to gradually tighten and circularize. By the time the orbit becomes circular the orbital separation shrinks to $a\sim5\,841\,\rsun$. But the tidal interactions are still at work trying to speed up the slowly rotating massive star and synchronize it with its companion which is spinning 3 orders of magnitude faster than the giant. The synchronization cannot be reached and Darwin instability sets in leading to further orbital decay until the giant star overflows its Roche lobe. Up to this moment the evolved star has lost $\sim57\%$ of its initial mass (its mass is now $17.2\,\msun$), the mass of MS star has remained almost unchanged and the orbital separation between the stars decreased to $a\sim3\,256\,\rsun$. The mass transfer that follows is dynamically unstable even if one considers the more restrictive conditions for unstable mass transfer than in our standard model (see \citet{Pavlovskii2017_CE,Olejak21} for details). At this point the evolved donor is a red supergiant with the convective envelope and low effective temperature $T_\mathrm{eff}\sim3.55\times10^3$ K. The sufficiently low binding energy of such an envelope leads to its successful ejection during the common envelope (CE) phase.\\
 The giant star loses its envelope becoming a Wolf-Rayet star on a close orbit ($a\sim140\,\rsun$) with MS star. In the subsequent evolution that lasts $\sim0.22$ Myr, the strong winds drive further mass loss from the primary star and the orbit expands to $a\sim164\,\rsun$ just prior the supernova (SN) explosion. The SN explodes $\sim5.7$ Myr from the onset of the binary evolution. The primary star becomes $9.7\,\msun$ BH and the effective kick (the composition of Blaauw and natal kicks) imparted to the system increases both its orbital separation and the eccentricity to $a\sim319\,\rsun$ and $e=0.49$ leading to the formation of Gaia~BH1-like binary.\\
Our synthetic Gaia~BH1 binary evolution proceeds on the nuclear timescale of the secondary which slowly evolves towards the end of hydrogen burning in its core. It takes $\sim5$ Gyr for the star to leave the main sequence and become a red giant what marks the end of Gaia~BH1 stage of the binary evolution. As the radius of the red giant expands the tides act to circularize the orbit. About $\sim600$ Myr later the red giant radius reaches the radius of the star Roche lobe and the mass transfer onto BH begins. The system turns into low-mass X-ray binary (LMXB) with the orbital period $P_\mathrm{orb}\sim140$ days ($a\sim251\,\rsun$) and donor mass $M_2=1.2\,\msun$. The mass transfer proceeds from less to more massive binary component, therefore, the orbit widens and so does the radius of the secondary star Roche lobe. When the Roche lobe extends to the point that the star fits inside it again, the mass transfer ceases. It happens when the orbital separation is $a\sim892$ \(\rsun\), $M_\mathrm{BH}\sim10.2\,\msun$ and $M_2\sim0.6\,\msun$. The LMXB phase lasts for $\sim4$ Myr. From there, the secondary star continues its nuclear evolution until it transforms into a carbon-oxygen white dwarf (WD) of mass $M_\mathrm{WD}=0.5\,\msun$. At the Hubble time $t_\mathrm{Hub}=13.7$ Gyr, the distance between the two compact objects in the system is $a\sim1244\,\rsun$.  For the rest of its life BH--WD binary slowly spirals in due to the gravitational radiation. The system will not merge within Hubble time.\\

To illustrate the evolution of Gaia~BH2 binary we choose the binary which starts  as the system consisting of massive MS star of mass $22.6\,\msun$  and MS star of mass $1.1\,\msun$ orbiting each other on the moderately eccentric ($e=0.66$) wide orbit ($a\sim7\,475\,\rsun$).
As in the case of Gaia~BH1, the system first evolves on the more massive (primary) star nuclear evolution time. At first the orbital separation increases due to the stellar wind from the primary but when the primary becomes a core helium burning giant and its radius reaches $R_2\sim520.4\,\rsun$ the tidal force starts to control the orbital evolution. At this point the orbital separation ($a\sim7\,795\,\rsun$) starts to shrink. It takes $\sim0.32$ Myr for the tidal force to circularize the orbit. While the orbit keeps shrinking the primary evolves onto the early asymptotic giant branch (AGB). The mass of the primary just before it explodes as SN has decreased to $M_1=18.5\,\msun$ and the orbital separation has contracted by over a half of its initial size, down to $a\sim3043\,\rsun$. The massive star becomes a $9.2\,\msun$ BH after $\sim9.53$ Myr from the start of the evolution. The large amount of mass ($\sim50\%$) that it loses in the explosion results in high eccentricity ($e=0.77$) that the orbit gains mainly due to the Blaauw kick but at the same time the systemic velocity remains low $v_{\mathrm{sys}}\sim32.8 \mathrm{km\,s^{-1}}$ due to the low natal kick velocity ($v_{\mathrm{kick}}\sim34.8 \mathrm{km\,s^{-1}}$). \\
The system consisting of BH--RG forms $\sim6.2$ Gyr after the onset of the binary evolution. While RG expands the tides act to tighten and circularize the orbit. When the orbital separation and the eccentricity decrease to $a\sim1175\,\rsun$ and $e=0.59$ respectively, they fall into the ranges which define the binary as Gaia~BH2-like (see Table~\ref{Table:Gaia-like param}). This phase of the binary evolution lasts only for $\sim2$ Myr. In next $\sim4$ Myr the eccentricity drops to zero and the orbital separation reaches its minimum, i.e. $a\sim758\,\rsun$. At this point the wind from the expanding RG comes into play: the orbit widens again, and BH accrets part of the matter lost by the evolved companion. By the time He is ignited in the core ($\sim6.7$ Myr after orbit circularization) the star mass drops to $M_2=0.75\,\msun$, BH mass slightly increases to $M_{\mathrm{BH}}=9.4\,\msun$ and orbit widens to $a\sim1165\,\rsun$. The exhaustion of He in the core occurs after $\sim112$ Myr when the companion becomes an early AGB star (E-AGB) and after next $\sim5.7$ Myr it evolves to become the thermally-pulsing AGB star (TP-AGB). The AGB stage, when the star loses roughly $25\%$ of its mass and the orbit widens by almost a factor of two reaching $a\sim2217\,\rsun$, is followed by the formation of carbon oxygen WD (CO WD) of mass $M_\mathrm{WD}=0.57\,\msun$ orbiting BH of mass $M_\mathrm{BH}=9.5\,\msun$ at a distance $a\sim2247\,\rsun$. 

\begin{figure}
\centering
\resizebox{\hsize}{!}
{\includegraphics[width=\textwidth,angle=270]{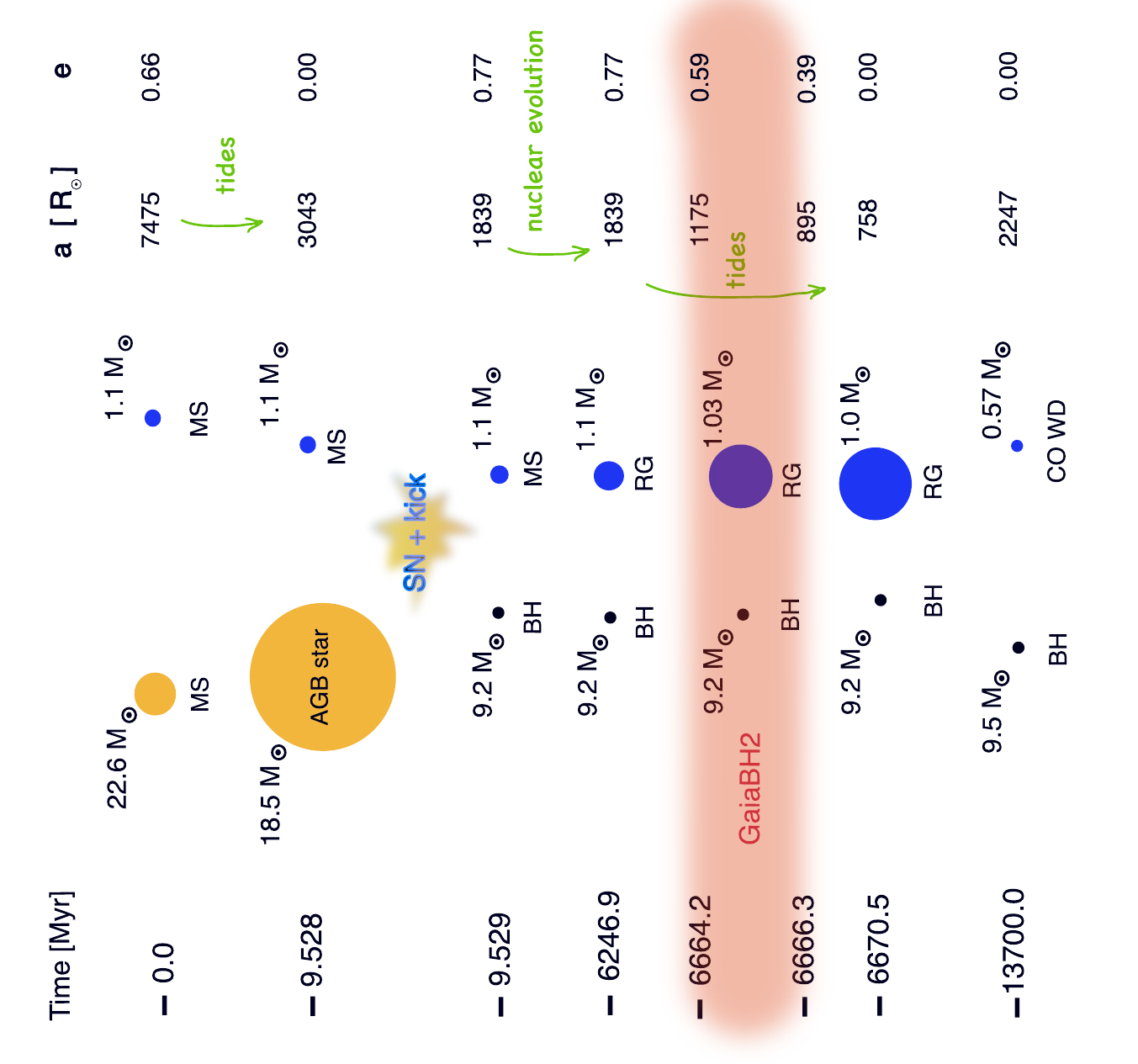}}
\caption{The main stages of Gaia~BH2-like binary evolution from the isolated binary evolution channel. The details are described in Sec.~\ref{ssec:results_evolution}.}
\label{Fig:GaiaBH2_evol}
\end{figure}

\subsubsection{Systemic velocities and natal kicks}\label{ssec:results_Vsys_Vkick}

\begin{figure}
\centering
\resizebox{\hsize}{!}
{\includegraphics[width=\textwidth]{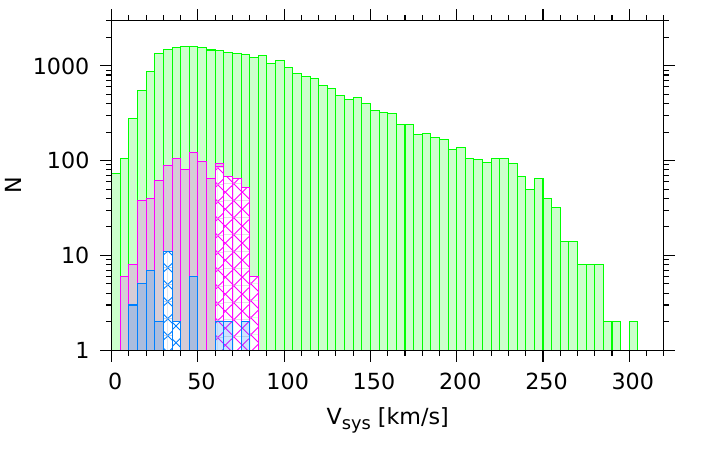}}
\resizebox{\hsize}{!}
{\includegraphics[width=\textwidth]{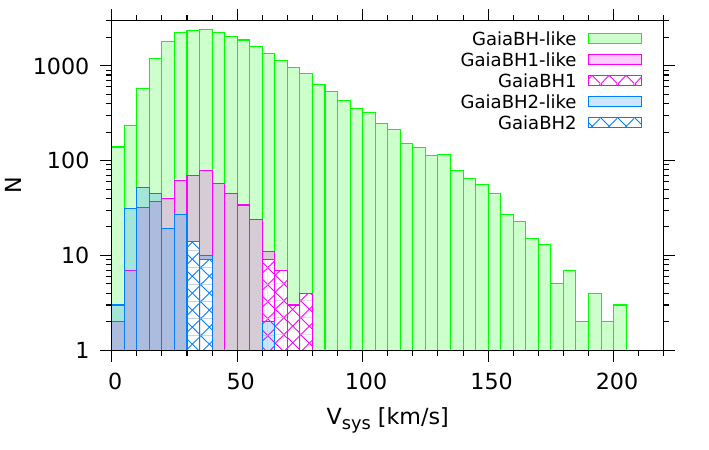}}
\caption{The distribution of the systemic velocity of all standard Gaia~BH-like binaries (green) in model V1 (top) and model V2 (bottom). The magenta distribution stands for the subset of Gaia~BH1-like binaries and the distribution in blue is for Gaia~BH2-like binaries. The binaries in both groups which have $v_\mathrm{sys}$ exactly in the observational ranges are shown as boxes with pattern. The criteria that we adopt for each Gaia~BH-like group are listed in Table~\ref{Table:Gaia-like param}.}
\label{Fig:Vsys_distr}
\end{figure}

\begin{figure}
\centering
\resizebox{\hsize}{!}
{\includegraphics[width=\textwidth]{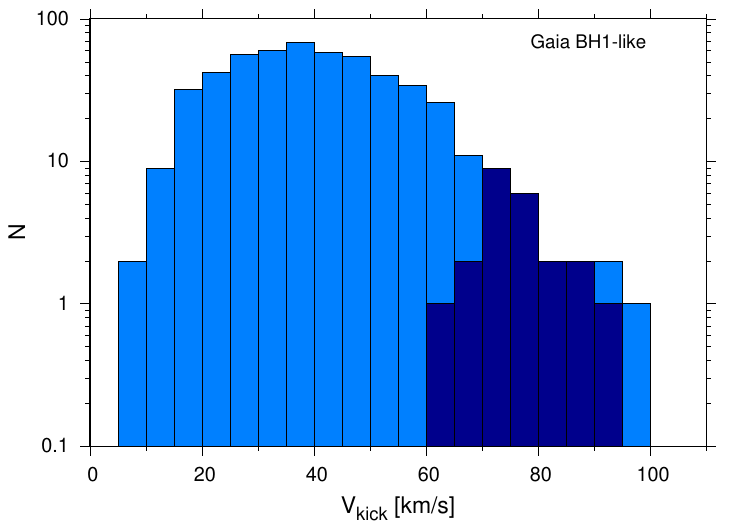}}
\resizebox{\hsize}{!}
{\includegraphics[width=\textwidth]{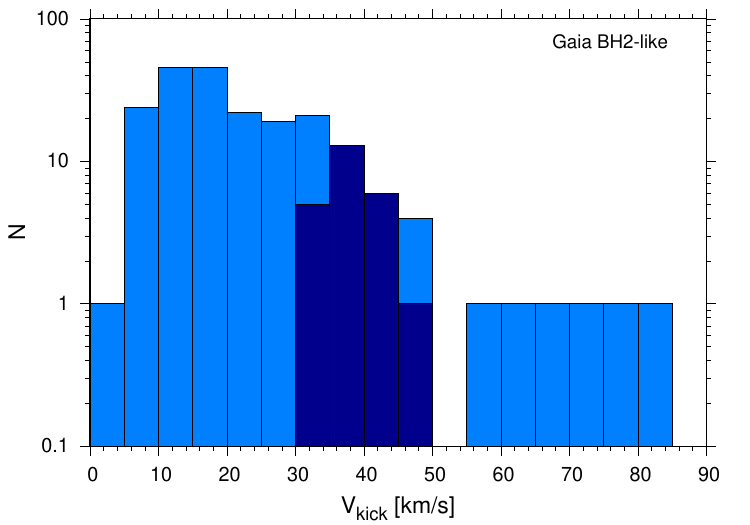}}
\caption{The distribution of the natal kick velocity received by the synthetic Gaia~BH1-like (top) and Gaia~BH2-like (bottom) in model V2. The dark-blue distribution stands for the subset of the binaries that have $v_\mathrm{sys}$ in the range corresponding to the observational estimations: $v_\mathrm{sys}\in[60.0-80.0]$ km/s for Gaia~BH1-like and $v_\mathrm{sys}\in[30.0-40.0]$ km/s for Gaia~BH2-like.}
\label{Fig:Vnk_distr}
\end{figure}

We present $v_\mathrm{sys}$ distributions of standard Gaia~BH-like binaries in models V1 (\citet{Hobbs05} natal kick velocity distribution) and V2 (\citet{Zhao2023} natal kick velocity distribution) in green in Fig.~\ref{Fig:Vsys_distr}. On top of them we show the distribution of $v_\mathrm{sys}$ for subsets of binaries which we identify as Gaia~BH1-like (magenta) and Gaia~BH2-like (blue), the systems which have $v_\mathrm{sys}$ exactly in the observational range for those systems are marked in pattern. 
The observed $v_\mathrm{sys}$ of Gaia~BH1 and Gaia~BH2 are within the ranges predicted by isolated binary evolution for Gaia~BH1-like and Gaia~BH2-like binaries. \\
For all standard Gaia~BH-like binaries the range of achievable $v_\mathrm{sys}$ is wide (up to $\sim300$ km/s in model V1 and $\sim200$ km/s in model V2) but the median values are well below $100$ km/s: it is $v_\mathrm{sys,med,V1}=72.8$ km/s for all Gaia~BH-like binaries in model V1 and $v_\mathrm{sys,med,V2}=45.3$ km/s in model V2. 
For Gaia~BH1-like binaries $v_\mathrm{sys,med,V1}\sim54.8$ km/s in model V1 and $v_\mathrm{sys,med,V2}\sim35.4$ km/s in model V2, for Gaia~BH2-like binaries $v_\mathrm{sys,med,V1}\sim19.2$ km/s in model V1 and $v_\mathrm{med,sys,V2}\sim16.7$ km/s in model V2. For Gaia~BH2-like binaries the median systemic velocity is comparable between the two natal kick velocity distributions while Gaia~BH1-like binaries move on average by $\sim55\%$ faster in model V1 than in model V2.\\
Overall there is $25\%$ more Gaia~BH-like binaries in model V2 than in model V1. Considering the two specific subsets of binaries it appears that the number of Gaia~BH1-like systems is almost the same in both models ($\pm3$ systems) but Gaia~BH2 are almost seven times more abundant in model V2. It is a straightforward consequence of the lower mean velocity of the natal kick velocities distribution of \citet{Zhao2023} and low velocities of natal kicks allow the wide and eccentric binaries like Gaia~BH2 to survive SN. \\
The corresponding median magnitude of the natal kick is not very different from $v_\mathrm{sys}$ due to the low mass of the companion star ($\sim1\,\msun$) and for standard Gaia~BH-like binaries it is $v_\mathrm{kick,med,V1}=82.9$ km/s in model V1 and  $v_\mathrm{kick,med,V2}=51.0$ km/s in model V2. \\
The distributions of the natal kick velocities received by Gaia~BH1-like and Gaia~BH2-like binaries for model V2 are shown on top and bottom panels of Fig.~\ref{Fig:Vnk_distr} correspondingly. The median natal kick magnitude is $v_\mathrm{kick}\sim39.3$ km/s for Gaia~BH1-like systems and $v_\mathrm{kick}\sim18.7$ km/s for Gaia~BH2-like systems. We mark the distributions of natal kick velocities for Gaia~BH1- and Gaia~BH2-like binaries which have the systemic velocities exactly in the observational limits \citep{Zhao2023} in dark-blue. \\
In our further analysis in Sec.~\ref{ssec:discussion_Vkick} we focus on model V2 as it is well motivated by observations, it gives us better statistics of Gaia~BH2 binaries and the conclusions of Sec.~\ref{ssec:discussion_Vkick} do not depend on the choice of $v_\mathrm{kick}$ distribution.

\subsection{Discussion}\label{sec:Discussion}

\subsubsection{The formation rate of Gaia~BH-like binaries}\label{ssec:discussion_formation_rates}

The total stellar mass of one IBE model is $M_\mathrm{sim}=3.917\times10^{11}\,\msun$. It comes from our choice to evolve only the binaries which may be the progenitors of Gaia~BH-like binaries, that is we take primary ZAMS mass to be $M_\mathrm{1,ZAMS}>18 \msun$ (so that it becomes BH) and the second star ZAMS mass to be $M_\mathrm{2,ZAMS}<1.5 \msun$. We then evolve $N=10^6$ such binaries.
To calculate $M_\mathrm{sim}$ we draw the binaries from the whole IMF and sum up their masses and we count the number of the subset of drawn binaries that have $M_\mathrm{1,ZAMS}>18 \msun$ and $M_\mathrm{2,ZAMS}<1.5 \msun$. We draw the binaries (summing up their masses) until we have a subset of N ($10^6$) binaries. We also take into account the median number of natal kick draws that are needed to get a Gaia~BH-like binary. Therefore, the calculated formation rates are corrected for using truncated IMF for both of the binary members and for the number of kick draws.\\ 
 If we scale down IBE simulations to match the total simulated initial cluster mass of star clusters simulations \mcltot$=3.6\times10^6\,\msun$ (see Sec.~\ref{res_clusters}) we find one Gaia~BH-like binary per simulation. This corresponds to the result of the dynamical interaction channel where also one Gaia~BH-like binary ejected from the cluster was found in a given model cluster set. In conclusion, both channels give the same formation rate that is $\sim2.7\times10^{-7}\,\msun^{-1}$.

\subsubsection{The estimated number of Gaia~BH-like binaries}\label{gaiabh_count}

A considerable amount of work has so far been done to estimate the number of dormant BHs that can be present in the Milky Way within \textit{Gaia} detection capability (e.g. \citet{Breivik2017,Mashian_Loeb2017,Yamaguchi2018,Wiktorowicz2020,Shikauchi2022,Chawla2022,Shikauchi2023}). Typically, the resulting number of binaries of interest varies from few tens to few hundreds \citep[][e.g.]{Yamaguchi2018,Wiktorowicz2020,Chawla2022}. However, the number can reach as high as $10^3-\sim10^5$ \citep[][e.g.]{Breivik2017,Mashian_Loeb2017} depending on the physics adopted in the evolutionary models, on the accuracy with which the Milky Way is modelled (where for example the realistic stellar distribution \citep[][e.g.]{Wiktorowicz2020,Chawla2022}), and/or how the position-dependent reddening \citep{Chawla2022} is taken into account.

Although the main focus of our paper is the comparison of the formation rates of Gaia~BH-like binaries for the isolated-binary and the dynamical formation channels, we complement this work with an approximate estimation of the expected number of Gaia~BH-like binaries in the Milky Way within the scope of \textit{Gaia} detectability, according to the two models. We limit our considerations to the Galactic thin disc population, since our simulations of the isolated binary evolution are done for a metallicity that is close to solar.

As an intrinsic synthetic population of Gaia~BH-like binaries, we choose that from the model ``V2 low-mass comp'' (see \ref{ssec:results_formation_rates}), where we assume that the natal kicks distribution is a double peaked Maxwellian and require the mass of the companion to be smaller than $3.0\,\msun$ for the binary to be classified as Gaia~BH-like.
We calibrate the mass of the simulations to match the stellar mass of the Galactic thin disc, where, for the latter, we adopt $M_\mathrm{d,thin}=2.15\times10^{10}\,\msun$ following \citet{Breivik2017} who assumed a constant star formation rate of $2.15\,\msun\mathrm{yr}^{-1}$ over 10 Gyr. We draw the present-day population of BH--low-mass star binaries from this intrinsic binary population, taking into account the estimated age of Gaia~BH1 ($\sim8$ Gyr) and the estimated duration of the red giant phase of the companion star in Gaia~BH2 ($\sim100$ Myr).

In the next step, the resulting binary population is spatially distributed at random, following the thin disc mass distribution described by the function from \citet{Yu_Jeffry2015}, i.e.,

\begin{equation}
\rho(R,z)=\frac{M_\mathrm{d,thin}}{4\pi h_R\,h_z}e^{R/h_R}\,sech^2(-z/h_z),
\end{equation}

where $h_R=2.5$ kpc is the scale lenght of the disc, $h_z=0.352$ kpc is its scale height, and $R$ and $z$ are the cylindrical coordinates of the axisymmetric disc.

\begin{figure}
\centering
\resizebox{\hsize}{!}
{\includegraphics[width=\textwidth]{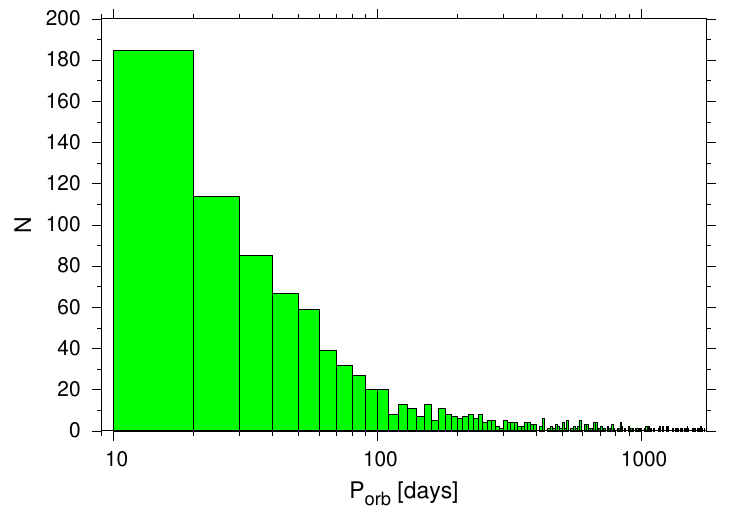}}
\resizebox{\hsize}{!}
{\includegraphics[width=\textwidth]{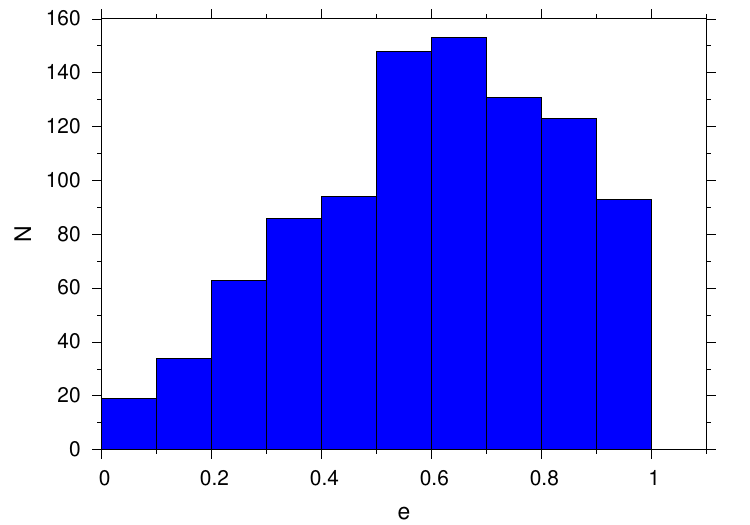}}
\caption{The distribution of the orbital periods $P_\mathrm{orb}$ (top) and eccentricity $e$ (bottom) of the predicted population of Gaia~BH-like binaries evolved through IBE detectable by \textit{Gaia} during 5 yr observational period.}
\label{Fig:Gaia_detect_IBE}
\end{figure}

To calculate the apparent magnitude $m_V$ of all considered present-day Gaia~BH-like binaries, we first calculate the absolute magnitude and the bolometric correction \citep{Torres2010} for each binary system in the sample using the luminosity, $L_2$, and the effective temperature, $T_\mathrm{eff,2}$, of the companion star. For the interstellar extinction in the $V$-band, $A_V$, we follow the approach of \citet{Yamaguchi2018} who assumed that $A_V(d)=d$. Here, $d$ in the distance from Earth in kpc that we calculate from the binary's position in the thin disc. We set a lower limit of $m_V<20$ mag for the apparent magnitude of the binary, if it is to be detected by \textit{Gaia}.\\
The astrometric signature, $\alpha$, is calculated following \citet{Breivik2017}:
\begin{equation}
    \alpha=\biggl(\frac{a_\mathrm{proj}}{\mathrm{au}}\biggr)\biggl(\frac{d}{\mathrm{pc}}\biggr)^{-1}\,\,\,\mathrm{arcsec}
\end{equation}
where $d$ is the distance to the binary and $a_\mathrm{proj}$ is the projected semi-major axis of the luminous companion. \\
We calculate $a_\mathrm{proj}$ for each system, using the exact position of the star on its orbit around the compact object and the orbital elements $i$, $\Omega$ and $\omega$ at a randomly chosen time elapsed from the star's periastron passage. We consider the binary as detectable if $\alpha>3\sigma_\mathrm{G}$, where $\sigma_\mathrm{G}$ is \textit{Gaia’s} astrometric precision calculated following \citet{Gaia_collab2016b} (see Eq.4 therein).\\
The final estimated number of BH--low-mass star binaries expected in the Galactic thin disc from the model of isolated binary evolution, assuming the most optimistic classification criteria, is of the order of $1.4\times10^3$ and, among them, $\sim900$ binaries are within \textit{Gaia}'s detectability limit. In Fig. \ref{Fig:Gaia_detect_IBE}, we present the distributions of the orbital periods and eccentricities of those binaries. It can be seen that binaries with orbital period shorter than $100.0$ days ($\sim66\%$ of all binaries) and eccentricity larger than $0.5$  ($\sim68\%$ of all binaries) dominate the distributions. About half of the binaries ($\sim44\%$) have both $\porb$ and $e$ in these ranges.

\subsubsection{The evolution Gaia~BH-like binaries}
In the following paragraphs we focus on the three phases of the binary evolution that we find important to discuss. The thorough studies of the impact of the choice of SN model (rapid vs delayed, with or without fallback), or of changing the value of the common envelope efficiency parameter $\alpha_\mathrm{ce}$ on the formation of detached BH--star binaries can be found for example in \citet{Chawla2022} and \citet{Shikauchi2023}. Although our results differ from conclusions of \citet{Shikauchi2023} on $\alpha_\mathrm{ce}$  (see Sec. \ref{ssec:discussion_lambda}) it is instructive to see the influence of $\alpha_\mathrm{ce}>1$ on the dormant BH populations.
\paragraph{The evolution Gaia~BH-like binaries: the impact of the natal kicks.}\label{ssec:discussion_Vkick}
\hfill \break

\begin{figure}
\centering
\resizebox{\hsize}{!}
{\includegraphics[width=\textwidth]{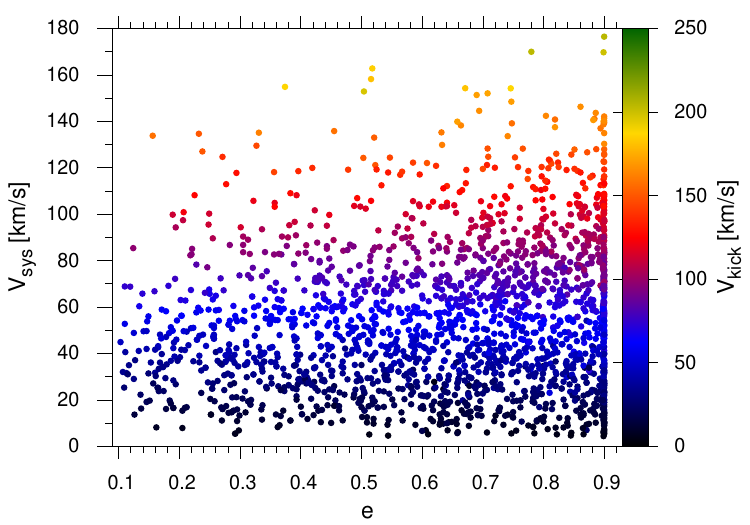}}
\caption{The relation between the binary eccentricity $e$, systemic velocity $v_\mathrm{sys}$ (km/s) and natal kick $v_\mathrm{kick}$ (km/s) magnitudes in model V2 for "optimistic" Gaia~BH-like binaries. We note that a given $v_\mathrm{kick}$ may result in a whole range of eccentricities.}
\label{Fig:Vkick2_wider_e_Vsys_Vkick_}
\end{figure}

\begin{figure}
\centering
\resizebox{\hsize}{!}
{\includegraphics[width=\textwidth]{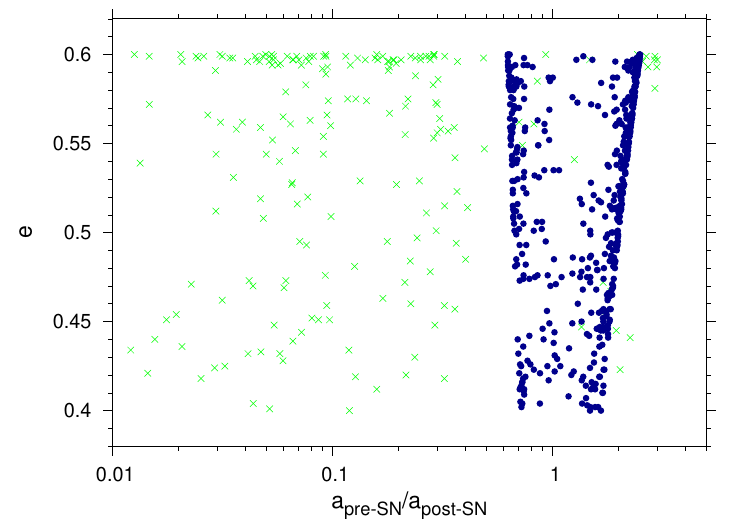}}
\caption{The relation between the change of the binary orbital separation after and just before SN ($a_\mathrm{post-SN}$/$a_\mathrm{pre-SN}$) and the eccentricity ($e$). The blue circles are Gaia~BH1-like binaries and the green crosses are Gaia~BH2-like binaries. For binaries which tend to form Gaia~BH1-like systems the fraction by which the orbit may change during first SN is strictly limited.}
\label{Fig:Vkick2_afai_e_BH1_BH2}\end{figure}

The natal kick velocity $v_\mathrm{kick}$ is a vector that plays an important if not the decisive role in the Gaia~BH-like systems formation.  What counts is not only the magnitude of the kick but also its direction and relative position of the stars on their mutual orbit when SN takes place.\\
The natal kick of a given magnitude determines $v_\mathrm{sys}$ but not the eccentricity of the binary, what is demonstrated in Fig.~\ref{Fig:Vkick2_wider_e_Vsys_Vkick_}. However, there is a correlation between the eccentricity, post- and pre-SN orbital separation \citep[see e.g.][]{Kalogera96} which is followed by our population of Gaia~BH-like binaries as is shown in Fig.~\ref{Fig:Vkick2_afai_e_BH1_BH2}. For Gaia~BH1, where the companion star is still on its main sequence, the evolution has not have enough time to change the binary orbit from its immediate post-SN configuration. The system may have widen or shrunken due to the natal kick but the total amount of the orbital separation change is tightly connected to the system post-SN eccentricity. If the natal kick causes binary orbit to widen the orbital separation may increase no more than twice ($a_\mathrm{post-SN}/a_\mathrm{pre-SN}\sim2$) for binaries that have eccentricities $e\in[0.4;0.6]$. If the orbit shrinks due to the natal kick then it can shrink by no more than $40\%$ ($0.6<a_\mathrm{post-SN}/a_\mathrm{pre-SN}<1.0$) (see Fig.~\ref{Fig:Vkick2_afai_e_BH1_BH2}) for such binaries. \\
There is no similar dependency for Gaia~BH2-like binaries. They harbor an already evolved star (RG) and the tidal forces between the binary components influence the orbital separation and eccentricity during Gaia~BH2-like phase. In the simulations we catch most of Gaia~BH2-like binaries when they reach the upper limits for $a$ and $e$ that we set for Gaia~BH2-like classification during their post-SN evolution. The relation between  $a_\mathrm{post-SN}$, $a_\mathrm{pre-SN}$ and $e$ for those systems is not straightforward.

\begin{figure}
\centering
\resizebox{\hsize}{!}
{\includegraphics[width=\textwidth]{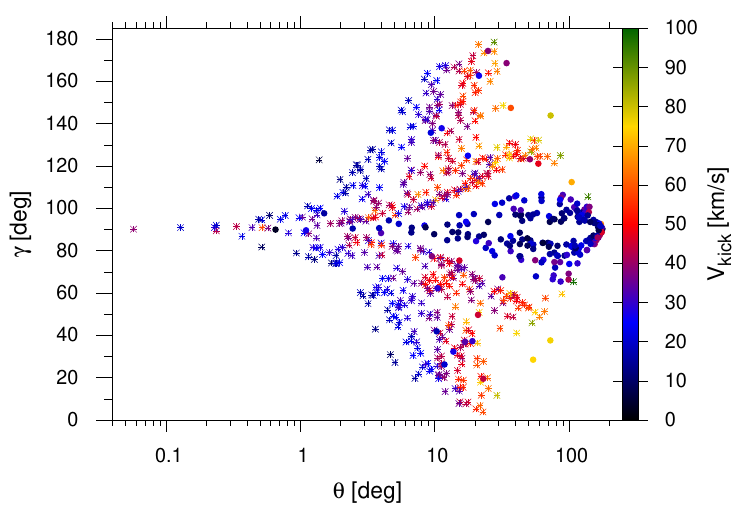}}
\resizebox{\hsize}{!}
{\includegraphics[width=\textwidth]{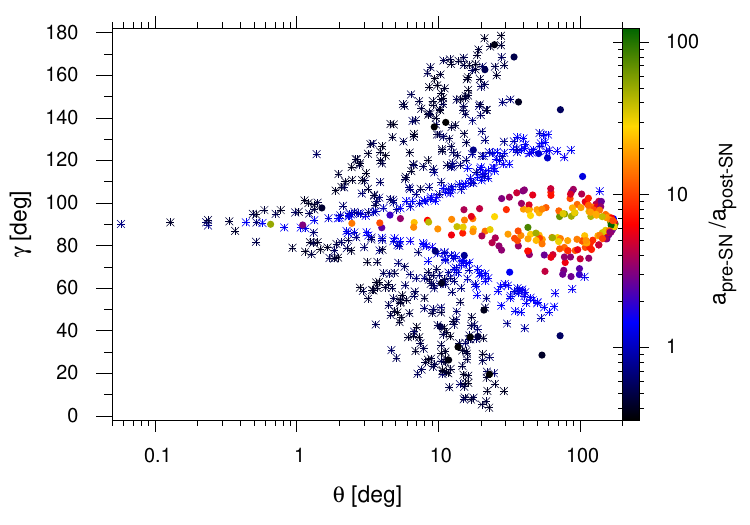}}
\caption{The relation between the misalignment angle between the binary angular momentum vector and BH spin vector after SN ($\theta$), the angle between the natal kick velocity vector and orbital angular momentum of the system $\gamma$ and natal kick velocity magnitude (top) or the the ratio of pre-SN to post-SN orbital separation (bottom). Both angles are in degrees units and $v_\mathrm{nk}$ is in km/s. Gaia~BH1-like systems are marked with stars and Gaia~BH2-like systems with filled circles. }
\label{Fig:Vkick2_teta_gamma_BH1_BH2}
\end{figure}

\begin{figure}
\centering
\resizebox{\hsize}{!}
{\includegraphics[width=\textwidth]{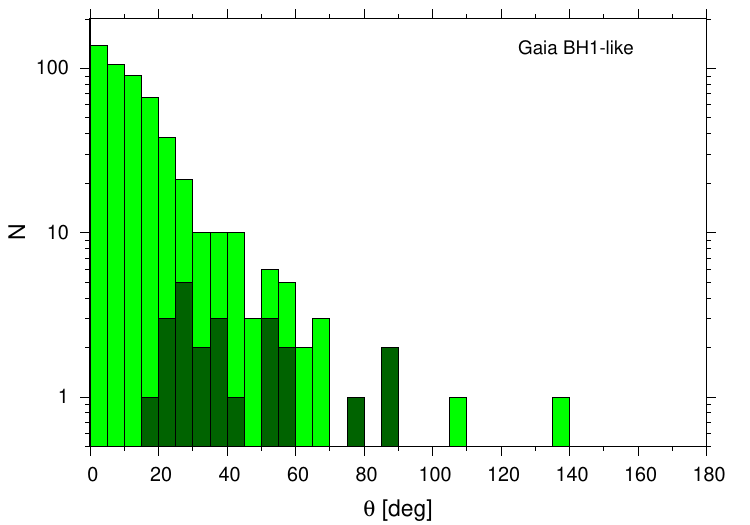}}
\resizebox{\hsize}{!}
{\includegraphics[width=\textwidth]{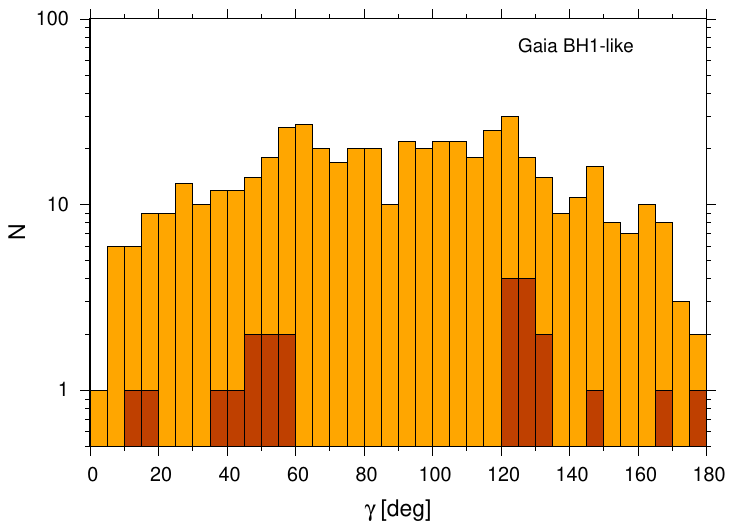}}
\caption{The distributions of $theta$ (top) and $\gamma$ (bottom) for Gaia~BH1-like systems. In dark color is marked the part of the distribution corresponding to Gaia~BH1-like binaries with $v_\mathrm{sys}$ in observational ranges.}
\label{Fig:Vkick2_teta_gamma_distrib_BH1}
\end{figure}

\begin{figure}
\centering
\resizebox{\hsize}{!}
{\includegraphics[width=\textwidth]{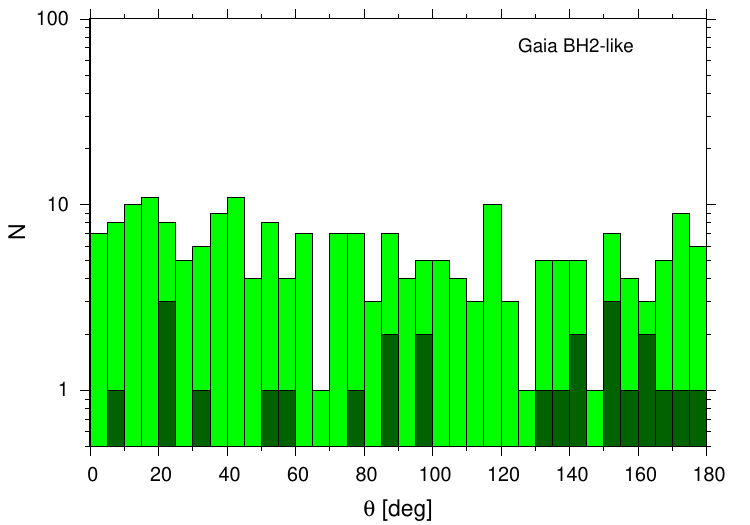}}
\resizebox{\hsize}{!}
{\includegraphics[width=\textwidth]{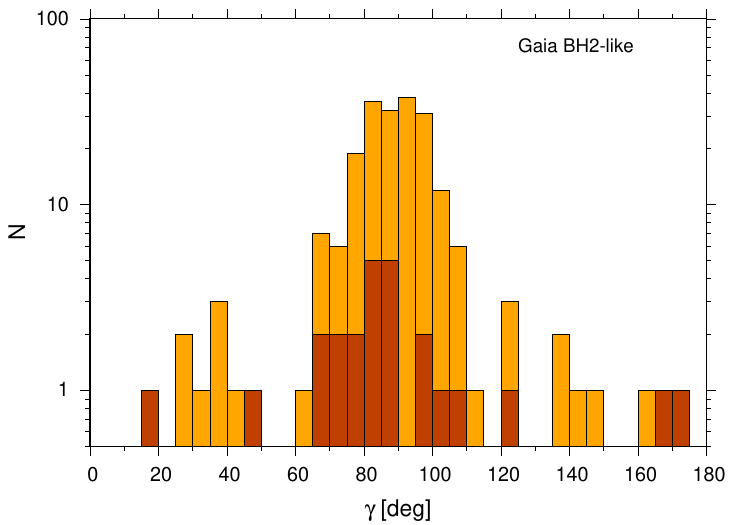}}
\caption{The distributions of $theta$ (top) and $\gamma$ (bottom) for Gaia~BH2-like systems. In dark color is marked the part of the distribution corresponding to Gaia~BH2-like binaries with $v_\mathrm{sys}$ in observational ranges.}
\label{Fig:Vkick2_teta_gamma_distrib_BH2}
\end{figure}

Whether the binary orbit shrinks or expands due to the natal kick depends on the position of the exploding star on the orbit around its companion at the moment of the SN explosion and on the natal kick vector, where both the position and the natal kick vector are drawn from the distributions described in Sec.~\ref{sec:IBE_Method}.\\
Apart from the shrinkage/widening of the orbit the natal kick changes the angle between the direction of BH spin and the direction of the binary angular momentum ($\theta$). We show the relation between angles $\theta$, $\gamma$ and $v_\mathrm{sys}$ (top) and $a_\mathrm{pre-SN}$/$a_\mathrm{post-SN}$ (bottom) for Gaia~BH1-like and Gaia~BH2-like binaries in Fig.~\ref{Fig:Vkick2_teta_gamma_BH1_BH2}. On both plots Gaia~BH1-like binaries are marked with stars and Gaia~BH2-like binaries are marked with filled circles. The formation of Gaia~BH1-like binaries depends differently on $\gamma$ than the formation of Gaia~BH2-like binaries.\\
Most of Gaia~BH1-like binaries form when the natal kick is directed out of the binary orbital plane. If $\gamma\sim90^{\circ}$ only a few ($<20$) Gaia~BH1-like binaries form and only with BH spin almost aligned with binary angular momentum vector ($\theta<2^{\circ}$). The higher is the natal kick velocity the further from the binary orbital plane the kick has to be oriented and the more BH spin is misaligned from binary angular momentum in Gaia~BH1-like binary, specifically $\sim94\%$ of Gaia~BH1-like population forms with $\theta<40^{\circ}$ and the median is $\theta_\mathrm{med}\sim10.5^{\circ}$ which can be seen in Fig.~\ref{Fig:Vkick2_teta_gamma_distrib_BH1}.\\
In the case of Gaia~BH2-like binaries $\gamma$ is a major factor determining their population. Most of them ($\sim95\%$) form if natal kick direction is close to the orbital plane ($\gamma\sim90^{\circ}\pm15$) and $v_\mathrm{kick}<41.6$ km/s. The binaries that survive SN and form Gaia~BH2-like systems have $\theta$ in the whole range of values. It can be clearly seen from the $\gamma$ and $\theta$ distributions shown in Fig.~\ref{Fig:Vkick2_teta_gamma_distrib_BH2}: $\theta$ distribution is almost flat and $\gamma$ distribution is centered around $90^{\circ}$. \\
There is a dependence between the natal kick direction and the tightening/expanding of binary orbit after SN. All systems marked in black in lower plot of Fig.~\ref{Fig:Vkick2_teta_gamma_BH1_BH2} have expanded due to the natal kick and the majority of them are Gaia~BH1-like binaries. The only Gaia~BH1-like binaries that have formed through tightening of the orbit after SN are those that form a pronounced narrow blue strip in bottom plot of Fig.~\ref{Fig:Vkick2_teta_gamma_BH1_BH2}. There is almost one-to-one relation between $\gamma$ and $\theta$ which allows the binary to shrink after SN and to form Gaia~BH1-like system.\\
The natal kick has to tighten the orbit if the binary is to form Gaia~BH2-like system. The wider is the pre-SN binary the lower velocity has to have the kick and the closer to the orbital plane it has to be oriented to become Gaia~BH2-like.

The distribution of angles $\theta$ and $\gamma$ for Gaia~BH1-like and Gaia~BH2-like binaries are presented on top and bottom plots in Fig.~\ref{Fig:Vkick2_teta_gamma_distrib_BH1} and Fig.~\ref{Fig:Vkick2_teta_gamma_distrib_BH2} respectively. The subsets of Gaia~BH1-like and Gaia~BH2-like binaries that have $v_\mathrm{sys}$ matching the observations are marked in dark colors. There is a specific range in $\gamma$ distribution which results in Gaia~BH1-like systems having observed $v_\mathrm{sys}$: most of them form if the natal kick acts at the angle $30^{\circ}-45^{\circ}$ to the orbital plane. In general the systems with observational $v_\mathrm{sys}$ fit into the distributions of $\theta$ and $\gamma$ for Gaia~BH1-like and Gaia~BH2-like populations. 

\paragraph{The evolution Gaia~BH-like binaries: the tidal interactions.}\label{sec:discussionIBE_tidal}
\hfill \break

To calculate the evolution of the orbital separation and eccentricity due to tidal interaction between the binary components, the convective damping in equilibrium tide and radiative damping in the dynamical tide \citep{Zahn77} are taken into account following the formalism of \citet{Hut1981}. The discrepancies between the tidal circularization periods derived from observations of coeval binary stars populations \citep{Meibom2005} and from synthetic binary populations have led to the introduction of an additional scaling factor, $F_\mathrm{tid}$, in the tidal evolution equations of orbital separation and eccentricity in $\st$. It appeared that to match the observed cut off period in the open cluster M67 with simulations, the scaling factor has to be set to $F_\mathrm{tid}=50$ in the case of convective damping \citep{Belczynski08}. Although different approaches has been proposed to determine the circularization periods of binary populations (e.g. \citealt{Zanazzi2022,Bashi2023}) and there are also new approaches to the theory of tides (e.g. \citealt{Terquem2021,Terquem_Martin2021}),  it seems that the need of enhancement of convective damping is maintained \citep{Mirouh2023}. \\
To evaluate the impact of $F_\mathrm{tid}$ on our results on Gaia~BH-like binary formation, we tested three values of the scaling factor: $F_\mathrm{tid}=1.0$, $50.0$ and $100.0$. We find that setting $F_\mathrm{tid}=1.0$ (the standard assumption in most binary population synthesis calculations) reduces the number of Gaia~BH-like binaries in the ``V2 standard'' model by a factor of $3.2$ (from $5550$ to $1757$ binaries), while $F_\mathrm{tid}=100$ increases that number only by $5\%$ (from $5550$ to $5815$). Changing the efficiency of tidal interaction in the convective case impacts mainly the binaries that we classify as Gaia~BH1-like: their number drops from $219$ for $F_\mathrm{tid}=100.0$ to $51$ for $F_\mathrm{tid}=1.0$. However, it has no such effect on Gaia~BH2-like binaries ($166$ binaries for $F_\mathrm{tid}=1.0$, $73$ for $F_\mathrm{tid}=50.0$ and $96$ for $F_\mathrm{tid}=100.0$).
The reason for the reduction in the number of Gaia~BH1-like binaries can be seen from their evolutionary path. At first, the orbit of the system widens due to the wind mass-loss from the massive primary star and, after a few Myr, the star evolves into the red supergiant phase when the radius becomes larger than $1000\,\rsun$ and the star's envelope becomes convective. This is when the tidal interaction between the stars sets in and the efficiency of the convective damping impacts the subsequent binary evolution. For $F_\mathrm{tid}=50.0$, the efficient convective damping causes the orbital separation to start shrinking and, at the point when the evolved star's radius reaches almost $2000\,\rsun$, the binary components are close enough to enter a CE phase. On the other hand, for $F_\mathrm{tid}=1.0$, the orbit does not shrink at all. The binary widens throughout the evolution of the massive primary due to the latter's stellar wind, until supernova explosion disrupts the binary. That way, the initial parameter space for the binaries to form Gaia~BH1-like system is much more limited for $F_\mathrm{tid}=1.0$.

There is no such effect in the case of Gaia~BH2-like binaries because they are much wider than Gaia~BH1-like binaries and most of them do not need to get close enough to evolve through a CE phase. The formation of Gaia~BH2-like binaries is mainly determined by the magnitude and the direction of the natal kick. The difference between the $F_\mathrm{tid}=1.0$ and the $F_\mathrm{tid}=100.0$ models is that most of the progenitors of Gaia~BH2-like binaries start their evolution with smaller orbital separations ($a_0=4000-7000\,\rsun$) in the former case than in the latter case ($a_0=6000-9000\,\rsun$).

\paragraph{The evolution Gaia~BH-like binaries: the common envelope parameters}\label{ssec:discussion_lambda}
\hfill \break

The results from the IBE evolution channel depend on the choice of two parameters that determine the CE evolution of the binary, namely, the envelope binding energy parameter $\lambda_{\mathrm{ce}}$ and the envelope efficiency parameter $\alpha_{\mathrm{ce}}$. In this section, we discuss the choice of $\lambda_{\mathrm{ce}}$ and $\alpha_{\mathrm{ce}}$ for our models.\\
As was mentioned in Sec.\ref{sec:IBE_Method}, for the calculations done for the purpose of this paper we adopt the $\lambda_{\mathrm{ce}}$ formulae by \citet{Wang2016} for stars more massive than $18.0\,\msun$, which we find to be more complete compared to the older prescriptions.The reason is that to calculate the binding energy of an envelope, \citet{Wang2016} take into account not only the contributions from the gravitational binding energy, internal energy of the envelope consisting of the thermal energy, radiation energy, and ionization and dissociation energies but also the enthalpy \citep[see][and discussion therein]{Ivanova_Chaichenets2011}. The inclusion of enthalpy decreases the envelope's binding energy with respect to the cases where it is not taken into account. This approach differs from the commonly used $\lambda_{\mathrm{ce}}$ prescription given by \citet{Claeys2014}, where only gravitation and ionization are considered as the energy sources that unbind the donor's envelope. To see how different choices of $\lambda_{\mathrm{ce}}$ impact the envelope binding energy for stars of different masses and metallicities, we refer to Fig.~A1 in the Appendix of \citet{Iorio23}, although it does not present the case of the \citet{Wang2016} formulae. From that figure, it can be seen that, for a $M_\mathrm{ZAMS}\sim30\,\msun$ star of metallicity $Z=0.01$, the envelope binding energy during the core He-burning phase is $\sim2$ orders of magnitude smaller for the $\lambda_{\mathrm{ce}}$ of \citet{Claeys2014} than for the $\lambda_{\mathrm{ce}}$ of \citet{Xu2010_Erratum}. Consequently, the envelope ejection during a CE phase is easier in the former than in the latter case, for the same value of $\alpha_{\mathrm{ce}}$ in both cases. This, in turn, may result in different post-CE orbital separations in models where, apart from $\lambda_{\mathrm{ce}}$, all other assumptions are identical.\\
The standard version of $\st$ has been using  $\lambda_{\mathrm{ce}}$ formulae from \citet{Xu2010,Xu2010_Erratum}. However, for the stars with ZAMS masses in the range of $\sim35-45\,\msun$, which are the progenitors of $\sim8-10\,\msun$ BHs, those formulae give $\lambda_{\mathrm{ce}}$ values as high as $4.0$ for the core He burning stars because they underestimate their envelope binding energy. The reason is that the formulae in \citet{Xu2010,Xu2010_Erratum} were fitted to the models calculated with the old Eddington's code EV only for stars with $M_\mathrm{ZAMS}<20\,\,\msun$. Those formulae were updated using Modules for Experiments in Stellar Astrophysics code (MESA, \citet{Paxton2011_MESA}) in 2016 by \citet{Wang2016} and should be used in their most recent form (Xiang-Dong Li private communication).
We further motivate our choice of the \citet{Wang2016} $\lambda_{\mathrm{ce}}$ by the fact that the formulae were fitted using a dense grid of stellar models calculated with MESA. Furthermore, they assumed a self-consistent core-envelope boundary located in the hydrogen-burning shell of the maximum compression prior to the CE, following \citet{Ivanova2011} \citep[see, however,][]{Klencki21}. The formulae from \citet{Wang2016}, that we use, were calculated using the prescription for the wind mass loss rates from \citet{Hurley00} and \citet{Vink2001} (Wind1 in \citet{Wang2016}) and we take $\lambda_{\mathrm{ce}}$ as the average of $\lambda_\mathrm{g}$ (corresponding to gravitational binding energy only), $\lambda_\mathrm{b}$ (corresponding to the total energy, comprising both gravitational and internal energy) and $\lambda_\mathrm{h}$ (which, in addition to the total energy, takes into account the enthalpy of the stellar envelope).\\

Note that our results are obtained with the standard value of $\alpha_\mathrm{ce}$, i.e., $\alpha_\mathrm{ce}=1.0$. There are analytical estimates for the final orbital separation, after the CE phase in pre-Gaia~BH1 systems \citep[e.g.][]{El-Badry23_GaiaBH1}. However, they are based on the assumption that the donor star has to lose a mass of $20\,\msun$ in envelope, during the CE. This is an overestimation, considering that the BH progenitor loses up to half of its mass in stellar winds during its evolution, before the CE phase starts (see Fig. \ref{Fig:GaiaBH1_evol} and its description in Sec.\ref{ssec:results_evolution}). Consequently, the envelope mass, that has to be unbound during the CE phase, has a mass of only $\sim3\,\msun$. If $M_\mathrm{env}=3.0\,\msun$ is considered in the calculations instead, the post-CE orbital separation of the binary turns out to be $>100\,\rsun$ instead of a few $\rsun$. \\
Taking into account the uncertainties of the $\lambda_{\mathrm{ce}}$ prescription and the correction to the post-CE orbital separation calculations as mentioned above, we conclude that invoking $\alpha_{\mathrm{ce}}>1$ should not be necessary to form Gaia~BH-like binaries via the IBE channel.
Also, it is so far unclear what (if any) physical mechanism could lead to $\alpha_{\mathrm{ce}}>1$. Therefore, we assume $\alpha_{\mathrm{ce}}=1$, which still seems to be an optimistic assumption as it implies that all orbital energy goes into unbinding the envelope without any energy loss.\\
For the analyses of the impact of different $\alpha_{\mathrm{ce}}$ values, of different SN engines (e.g., rapid and delayed), and of the reduction of SN natal kicks due to matter fallback on the isolated-binary formation of BH--star binaries, we refer to the works of \citet{Yamaguchi2018} and \citet{Chawla2022}.

\subsubsection{Gaia~BH1 as Low Mass X-ray Binary}\label{ssec:discussion_LMXB}

There is no X-ray emission detected from Gaia~BH1 and Gaia~BH2. However, it is found that during the post-dormant BH binary phase, both of these binaries may evolve through the symbiotic X-ray binary phase to the phase of stable Roche-lobe overflow (RLOF) \citet{Rodriguez2023}. At this point, the systems convert into low mass X-ray binaries (LMXBs). The population of LMXBs, that could arise from Gaia~BH-like systems, would have long orbital periods of $P_\mathrm{orb}>100$ days.\\
Of the two binaries that we consider as the best examples of Gaia~BH1 and Gaia~BH2 in our IBE simulations and which evolution we describe in Sec.~\ref{ssec:results_evolution} only Gaia~BH1-like system undergoes RLOF becoming LMXB in its subsequent evolution while Gaia~BH2-like binary remains detached.\\
As of today, all known LMXBs with BH accretors are transient sources, that is they show periodic rapid increase in their luminosity up to $L_\mathrm{X}\sim10^{39}$ erg/s lasting from days to years, followed by long periods (tens of years) of quiescence when their luminosity drops below $L_\mathrm{X}\sim10^{33}$ erg/s. The outbursts are triggered by an instability in accretion disk formed around BH. The instability arises from the changes in opacities when ionisation/recombination sets in, what in turn entails the changes in the disk viscosity (for details of disk instability model (DIM) see e.g. \citet{Lasota01,Hameury20}). According to the model the system is persistently bright (has high level of luminosity with no outbursts) if the mass transfer rate $\dot{M}_\mathrm{tr}$ is higher than critical value of mass accretion rate at the outer disk radius ($\dot{M}^{+}_\mathrm{crit}(R_\mathrm{d})$) and the system is transient (undergoes outbursts) if $\dot{M}_\mathrm{tr}<\dot{M}^{+}_\mathrm{crit}(R_\mathrm{d})$ and $\dot{M}_\mathrm{tr}>\dot{M}^{-}_\mathrm{crit}(R_\mathrm{in})$ where $\dot{M}^{-}_\mathrm{crit}(R_\mathrm{in})$ is the critical mass accretion rate at the inner disk radius. The formulae for $\dot{M}_\mathrm{crit}(R)$ are taken from \citet{Lasota08}:
\begin{equation}
\begin{split}
    & \dot{M}^{+}_\mathrm{crit}(R)=8.07\times10^{15}\alpha_{0.1}^{-0.01}\,R_{10}^{2.64}\,M_1^{-0.89}\,& \mathrm{g\,s^{-1}} \\
    & \dot{M}^{-}_\mathrm{crit}(R)=2.64\times10^{15}\alpha_{0.1}^{0.01}\,R_{10}^{2.58}\,M_1^{-0.85}\,&\mathrm{g\,s^{-1}}
\end{split}
\end{equation}
where $\alpha_{0.1}=0.1\alpha$ is the viscosity parameter in units $0.1$, $R_{10}$ is the radius in units of $10^{10}$ cm and $M_1$ is the primary mass in solar masses.\\
In following calculations we assumed that the disk outer radius extends to $90\%$ of BH Roche lobe radius in the outburst and that the inner disk radius $R_\mathrm{in}$ does not coincide with the last stable orbit in the quiescence but instead it is truncated due to the transition from optically thick to optically thin radiatively inefficient flow \citep[e.g.][]{Lasota96,DeMarco21}. We further assumed that the truncation happens at $3.4\times10^4 R_\mathrm{g}$ where $R_\mathrm{g}$ is gravitational radius $R_\mathrm{g}=GM_\mathrm{BH}/c^2$. $\dot{M}^{-}_\mathrm{crit}(R)$ decreases with $R$ and the assumption about $R_\mathrm{in}$ is the upper limit for the inner disk truncation that is justified by the observations of V404 Cyg (a transient LMXB) \citep{Bernardini16}. The viscosity parameter in the hot disk is taken to be $\alpha_{0.1}=1.0$ and in the cold disk $\alpha_{0.1}=0.2$ \citep{Lasota01}. We calculated the critical mass accretion rates at the outer and inner disk radii for parameters that we found for exemplary Gaia~BH1-like systems described in Sec.~\ref{ssec:results_evolution} at the onset of LMXB phase:
\begin{equation}
\begin{split}
    & \dot{M}^{+}_\mathrm{crit}(R_\mathrm{d})=7.95\times10^{-6}\, & \mathrm{M_\odot\,yr^{-1}} \\
    & \dot{M}^{-}_\mathrm{crit}(R_\mathrm{in})=3.93\times10^{-10}\,& \mathrm{M_\odot\,yr^{-1}}
\end{split}
\end{equation}
for $R_\mathrm{d}=20.44$ \(\rsun\) and $R_\mathrm{in}=0.71$ \(\rsun\).\\
The mass transfer rate for the considered binary in its LMXB state is $\sim3.53\times10^{-8}\,\msun\,\mathrm{yr}^{-1}$, therefore, according to the disk instability model, the system should become a transient LMXB. If LMXBs that formed from Gaia~BH1-like binaries exist, they would be observed through their outbursts. It is of interest to predict what would be the outburst recurrence time of such systems.
The majority of the disk mass is accreted onto the compact object during the long outburst. The disk depleted from matter needs to refill so that another outburst may develop. Following \citet{Hameury_Lasota2020_eq} the refilling time corresponding to the outburst recurrence time is:
\begin{equation}
    t_\mathrm{recc}=\frac{\Delta M}{\Dot{M}_\mathrm{tr}},
    \label{eq:t_fill_1}
\end{equation}
where $\Delta M$ is the amount of the disk mass accreted during outburst and $\Dot{M}_\mathrm{tr}$ is the mass transfer rate.\\
We assumed that the amount of mass retained in the disk at the end of outburst is $10\%$ of the maximum disk mass (just before the outburst is triggered) and for the purpose of this calculations we took $\Delta M=0.9M_\mathrm{d}$, where the disk mass is:
\begin{equation}
M_\mathrm{d}=3.3\times10^{26}\psi\alpha_{0.2}^{-0.8}\Dot{M}_{19}^{0.7}M_1^{0.25}R_{12}^{1.25}\,\,\mathrm{g}
\label{eq:M_disc}
\end{equation}
where $\psi$ is a factor connected to opacity (for details see \citet{Hameury_Lasota2020_eq}) and we used $\psi=1.3$ following the authors, $\alpha$ is the disk viscosity parameter in units 0.2 \citep{Lasota01}, $\Dot{M}$ is the accretion rate in units of $10^{19}\,\mathrm{g/s}$, $M_1$ is the compact object mass in solar units and $R_{12}$ is the outer disk radius in outburst in units of $10^{12}$ cm.\\
The mass transfer rate (here equal to mass accretion rate) \footnote{The mass accretion rate on compact object is assumed to be equal to the mass transfer rate during the stable Roche lobe overflow in $\st$, but in general this is not true for the discs of transient LMXBs. However, it is a sufficiently good approximation for our purpose of the recurrence time estimation.} changes in time with the nuclear evolution of the donor and so does the Roche lobe radius of the accretor. Therefore, we estimate two limiting values of $t_\mathrm{recc}$: at the onset (minimum $t_\mathrm{recc}$) and at the end (maximum $t_\mathrm{recc}$) of LMXB phase of the considered Gaia~BH1-like system. The parameters used for calculations are give in Table~\ref{table:app_trecc} in Appendix. The obtained recurrence times are $\sim47.3$ yr and $\sim159.8$ yr respectively.
This suggests that there may be a population of long orbital period transient LMXBs that originated from dormant BH binaries and that are hard to observe because their luminosity rises above the detection threshold only once every few tens of years.

\section{Gaia BH binaries from young massive and open star clusters}\label{GaiaBH_clusters}

\subsection{Computed model clusters}\label{cluster_models}

It would be worth considering and comparing our isolated-binary formation rate of Gaia BH1- and Gaia BH2-like
binaries with those from star clusters. Dynamical interactions among stars, stellar binaries, and BHs in star clusters,
particularly in low mass open clusters, is widely proposed to be a channel for assembling detached
BH-star binaries \citep{Banerjee_2018,El-Badry23_GaiaBH1,Tanikawa2023,Rastello23_gaiaBH1_form,DiCarlo23_GaiaBH_form}.
In the present work, we consider the evolutionary model star cluster set of \citet[][hereafter Ba21]{Banerjee_2020c}
to estimate the formation rate of dynamical BH-star binaries in the galactic field.

We waive a detailed description and discussion of the computed models in Ba21, which are available in
the Ba21 paper itself and are discussed further in several follow-up studies
\citep[e.g.,][]{Banerjee_2020d,Banerjee_2021,Belczynski_2022}. To summarize, the Ba21 model set
comprise 72 model star clusters that evolve from a young massive phase to an old, open cluster-like
phase. The clusters' initial mass cover the range of $10^4\msun\leq\mcl\leq10^5\msun$ and initial size (half mass
radius) range between $1-3$ pc. About half of the clusters have an initial \emph{overall} primordial binary fraction between
$5\%-10\%$; however, the clusters' massive-star members, above $m_\ast > 16\msun$, 
are paired separately among themselves with an initial binary fraction of 100\%.
The clusters cover the metallicity range $0.0001 \leq Z \leq 0.02$.
All models are subjected to a solar-neighbourhood-like galactic tidal field.

The cluster models are computed with the direct, star-by-star N-body integration code
{\tt NBODY7} \citep{Aarseth_2012}, with various updates as described in Ba21.
In particular, stellar and binary evolution and remnant formation take place in the models
via the
coupling of {\tt NBODY7} with an updated version of the fast binary evolution code
{\tt BSE} \citep{Hurley_2002,Banerjee_2020}.
In most models, stellar remnants (NSs and BHs) form via the ``rapid'' remnant mass
model of \citet{Fryer12}, as implemented in \citet{Banerjee_2020}.

As seen in the above mentioned followup works of Ba21, these cluster models produce GW mergers
with properties and rate consistent with those inferred from GW observations.
The main motivation for choosing the Ba21 models for this study is that they are evolved
either at least until all the in-cluster BHs
are practically depleted or until $\approx11$ Gyr (see Appendix C of Ba21). Therefore, the population of
escaped BH-star binaries (and other escaped BH-containing systems) from these clusters can be
considered to be the complete yield, allowing for robust rate estimates for
such field objects from clusters.  
This is unlike any other recent study addressing the dynamical formation of detached
BH-star binaries (see above). Therefore, in the following,
we shall focus on only those BH-star binaries that have genuinely escaped from the
clusters, i.e., those which have crossed their parent cluster's tidal radius and have become
tidal-tail members. This would allow for a fair comparison with the yield from
isolated binaries (see above) and with Gaia BH1 and Gaia BH2, both of which
are field objects.

\subsection{Results: formation rate of field BH-star binaries}\label{res_clusters}

\begin{figure*}
\includegraphics[width=18.0cm,angle=0]{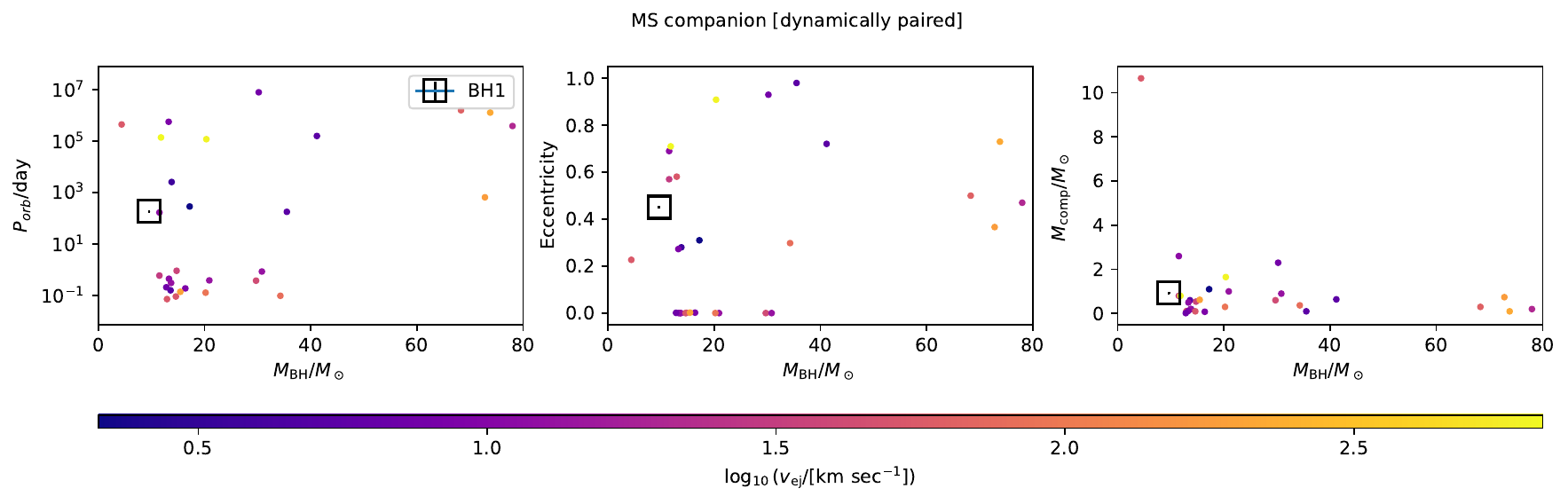}\\
\vspace{-1.2cm}
\includegraphics[width=18.0cm,angle=0]{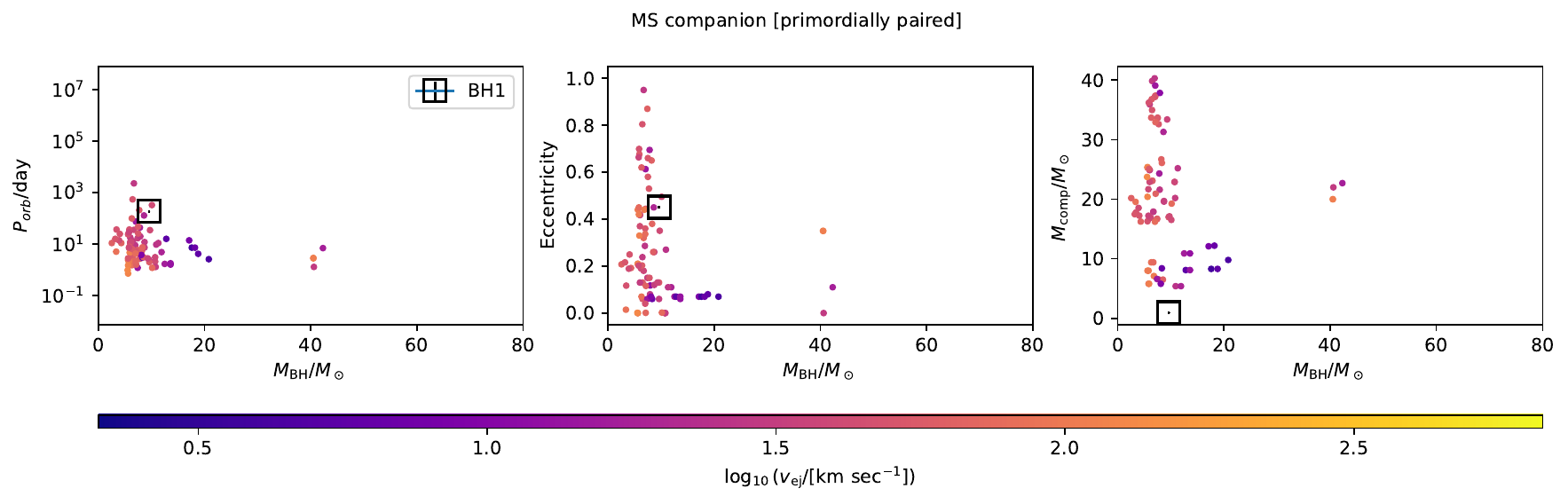}\\
\vspace{-1.2cm}
\includegraphics[width=18.0cm,angle=0]{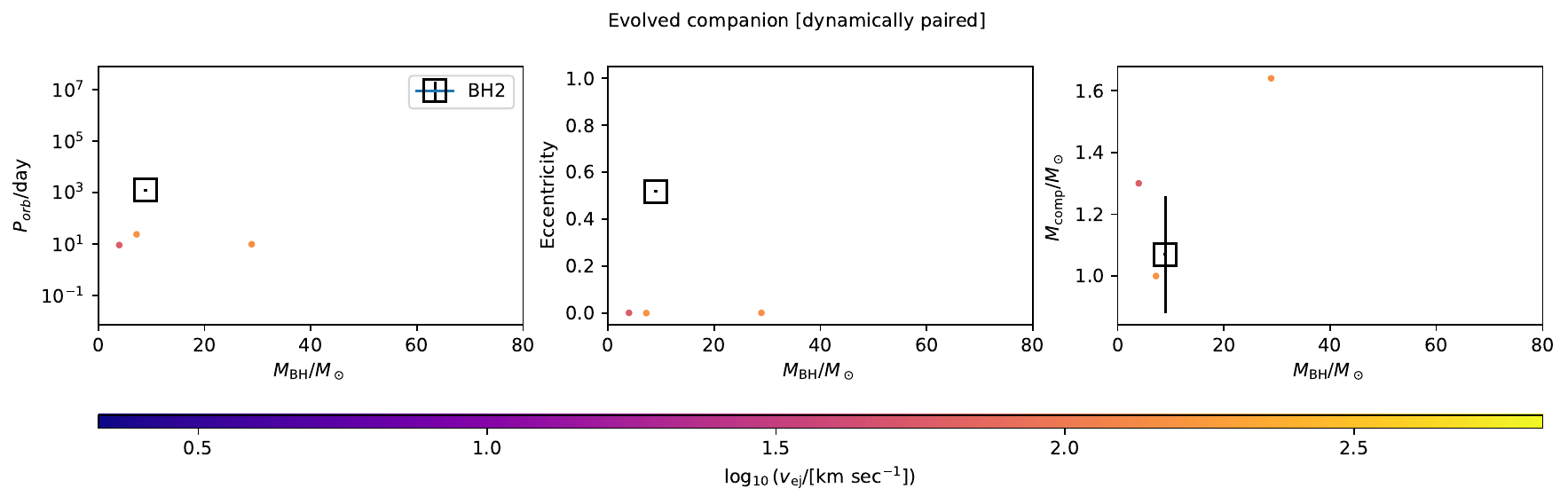}\\
\vspace{-1.2cm}
\includegraphics[width=18.0cm,angle=0]{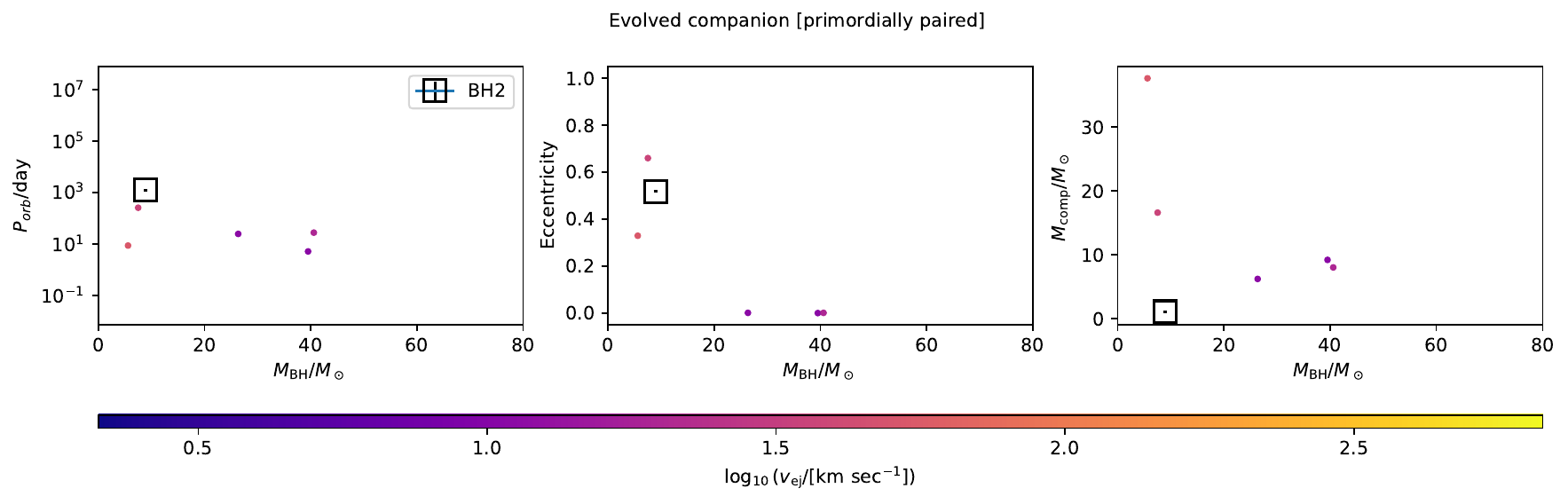}
\caption{Demographics of the BH-star binaries that have escaped into the galactic field
from the evolutionary model star clusters of Ba21. The filled circles in the panels show the mass of the
BH member ($\mbh$; X-axes) against the orbital period ($\porb$; left column),
eccentricity (middle column), and companion-star mass ($\mcomp$; right column)
of these model BH-star binaries. The binaries are plotted separately, based
on the evolutionary stage of the companion star (MS or evolved) and the assembly
channel (primordial or dynamical), as indicated in each row's title.
The data points are colour-coded according to the velocity of ejection
of the binary ($\vej$; colour bar). All of the plotted values correspond to the event
of the BH-star binary crossing the instantaneous tidal
radius of its parent cluster. For comparison, the observed BH-star binary candidates
Gaia BH1, that contains a main-sequence companion, and Gaia BH2, that
contains an evolved companion, are shown in the upper and lower pair of rows, respectively
(empty squares). For these observed data points, the colour coding is not followed
and their error bars are generally invisible due to the scale of the figure axes. (The size
of the squares are chosen for legibility and does not represent
measurement uncertainties.)
}
\label{fig:Ba21_scatter}
\end{figure*}

Fig.~\ref{fig:Ba21_scatter} shows the orbital period, $\porb$, eccentricity, $e$, and companion mass,
$\mcomp$, against BH mass, $\mbh$, of all the BH-star binaries that have escaped 
into the galactic field from the model clusters of Ba21. In all panels, the individual data points are
colour coded according to the velocity of ejection, $\vej$, which is the velocity
of the escaping binary's center of mass, at ejection. Unless otherwise stated, all quoted or plotted values
of a BH-star binary are those when the escaping binary crosses the instantaneous tidal radius
of its parent cluster; i.e., at the moment of its `formal' escape form the cluster.
These confirmed escaped binaries and their parameters are recovered from
the {\tt NBODY7} simulation logs, with the help of the ``BINARY ESCAPE'' records.
In Fig.~\ref{fig:Ba21_scatter}, the four rows distinguish between assembly mechanism
(dynamically or primordially assembled) and the stellar companion's evolutionary status
(main sequence or evolved) of the binary.

The figure demonstrates that escaped BH-star binaries with a low mass companion ($\mcomp\lesssim3\msun$)
are most likely dynamically assembled, as anticipated by \citet{El-Badry23_GaiaBH1}.
In the Ba21 clusters, such pairing between a low mass star and a BH can happen
only via dynamical channel since, initially, massive, BH-progenitor stars are always primordially paired
among themselves (and never with a low mass star). The dynamically assembled BH-MS binaries
generally cover the $\porb$, eccentricity, $\mcomp$, and peculiar velocity range of Gaia BH1. 
However, the escaped BH-evolved binaries (both dynamically and primordially paired)
struggle to meet the observed properties of Gaia BH2. As such, there are only a handful of
BH-evolved binaries that have escaped from the entire Ba21 set.

If all orbital parameters
are compared together, there is only one ejected BH-star binary that is somewhat
comparable to Gaia BH1: this binary has $\mcomp=1.1\msun$ (main sequence), $\mbh=17.2\msun$,
$\porb=290$ day, $e=0.31$. The total computed initial star cluster mass of 
the Ba21 set being $\mcltot=3.6\times10^6\msun$, this gives an unweighted (plain)
formation rate of Gaia BH1-like system, per unit initial cluster mass, of $2.7\times10^{-7}\msun^{-1}$.
This particular binary is formed via a dynamical exchange encounter in one of the
$\mcl=2\times10^4\msun$ cluster and is also ejected dynamically from the cluster.
In the Ba21 set, there is no ejected BH-star system that is or can evolve (after the ejection)
anywhere similar to Gaia BH2.

\begin{figure*}
\includegraphics[width=17.0cm,angle=0]{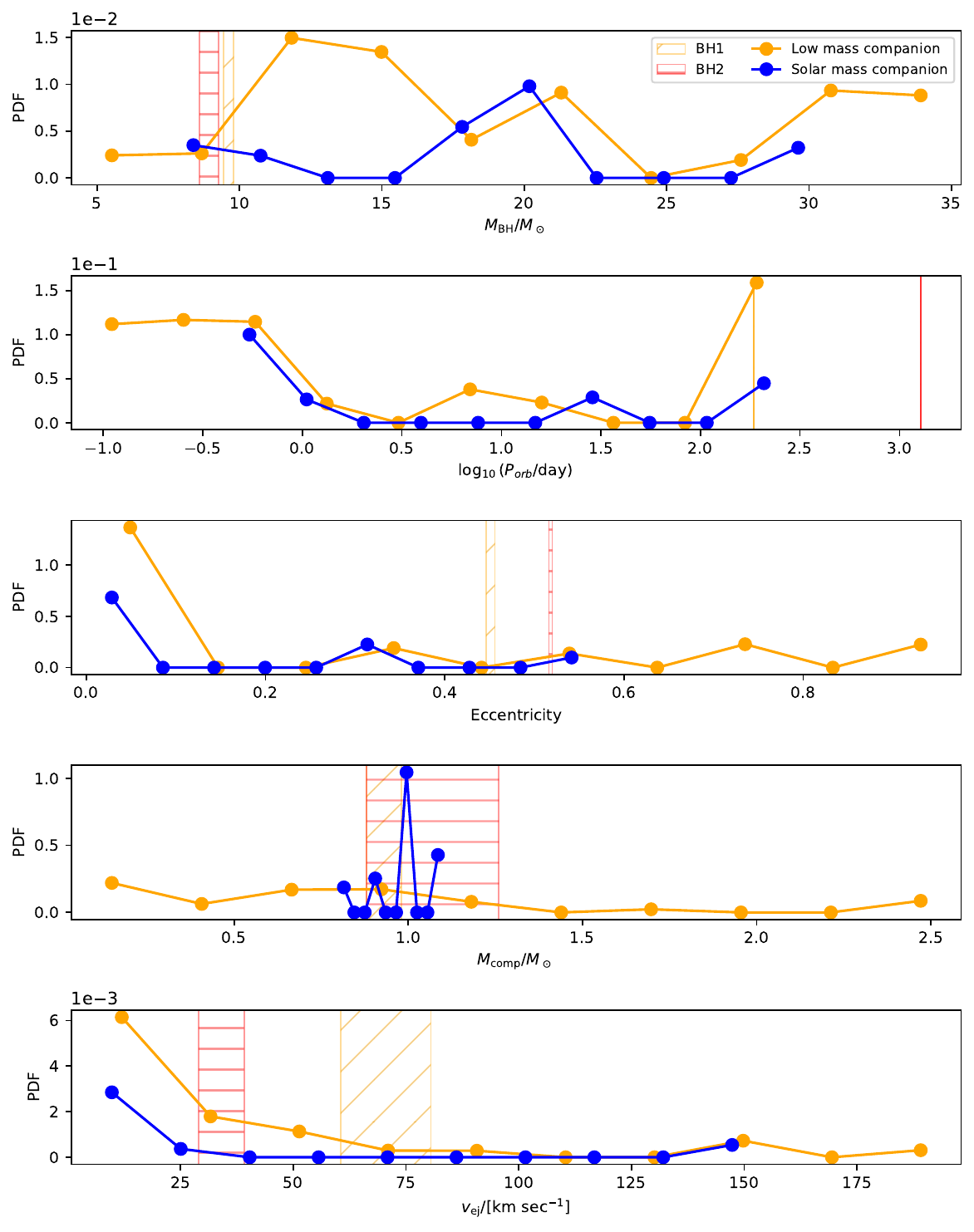}
\caption{Normalized distributions of BH mass ($\mbh$), orbital period ($\porb$), eccentricity,
companion star mass ($\mcomp$), and ejection velocity ($\vej$)
of BH-star binaries that have escaped into the galactic field from the evolutionary model star clusters of Ba21
(upper to lower panels; all values correspond to the event
of the BH-star binary crossing the instantaneous tidal radius of its parent cluster).
The distributions of binaries with a low mass companion ($\mcomp<=3\msun$; orange line) and with a solar
mass companion ($0.8\msun<=\mcomp<=1.1\msun$; blue line) are shown separately. These distributions
incorporate a weighted distribution of cluster birth mass, $\mcl$, according to the power law
$\propto\mcl^{-2}$. Furthermore, the two distributions on each panel are normalized according
to the relative number of BH-low mass and BH-solar mass binaries ejected from the weighted
cluster population. The observed Gaia BH1 and Gaia BH2 binaries are indicated in the panels;
in the final panel, peculiar velocities of the Gaia BH binaries are indicated.}
\label{fig:Ba21_dists}
\end{figure*}

Table~\ref{tab:rates_clusters} (middle column) shows the plain formation rates for escaped BH-star
systems with a low mass companion ($\mcomp\leq3\msun$), with a solar mass companion
($0.8\msun\leq\mcomp\leq1.1\msun$), and of Gaia BH1-type, as obtained from the Ba21 set.    
The table (right column) also states the corresponding weighted rates when the initial
masses of the clusters are assumed to follow a power-law distribution of index $-2$,
as motivated by observations \citep[e.g.,][]{PortegiesZwart_2010,Bastian_2012}.
The weighted rates are similar to the plain rates since there are only a handful
of field low-mass-companion systems (21, 5, and 1 for the three cases, respectively)
from a number of model clusters spanning over a wide initial mass range (see Sec.~\ref{cluster_models}).
Fig.~\ref{fig:Ba21_dists} shows the mass, orbital parameter, and
$\vej$ distributions of the escaped BH-star binary population in Ba21,
with low mass (orange line and circles) and solar mass
(blue line and circles) companion. The distributions are constructed assuming
a $\propto\mcl^{-2}$ initial cluster mass distribution. 
Note that there is only one specimen that is comparable to a Gaia-detected BH-star
binary (see above). The distributions are nevertheless presented keeping in mind
the overall diverse demographics of field BH-star binary population from these computed star
clusters (c.f. Fig.~\ref{fig:Ba21_scatter}), that may address such binaries
discovered in the future.

As evident from Figs.~\ref{fig:Ba21_scatter} and \ref{fig:Ba21_dists}, the vast
majority of dynamically assembled field BH-star systems have $\mcomp<3\msun$.
Nearly half of them have zero or small eccentricity, an MS companion, and $\porb<1$ day --- these
very tight binaries are outcomes of tidal capture between the BH and the companion
star inside the parent cluster and the binary's subsequent tidal circularization\footnote{In the Ba21
runs, tidal energy loss and analytical tidal circularization
were enabled (option $27=1$; \citealt{Aarseth_2003}).}. Inside clusters, such close binaries behave like
single, massive objects most of the time and are ejected dynamically in ways
similar to those of ejection of single BHs \citep{Kulkarni_1993}. These systems are potentially tidally interacting or
mass transferring and are not compatible with the Gaia detected, detached BH-star binaries.
There are also a few wide, $\porb\gtrsim 100$ day, MS-companion binaries --- these systems
are assembled in the cluster via dynamical exchange encounters and are also ejected
via dynamical scatterings (see above), making them eccentric in general.

Inside a cluster, dynamical formation of wide and eccentric BH-star binaries with a low mass
stellar companion is rather common \citep{Banerjee_2018}. After formation, such a binary would undergo
dynamical hardening whilst inside the cluster. However, the BH-star pair is also vulnerable
inside a cluster containing a dynamically active BH subsystem, since the low mass stellar
companion can easily be displaced by a much more massive BH via a close exchange encounter.
In other words, the in-cluster BH-star phase is at best a short-lived, intermediate phase
\citep{Banerjee_2017b}. This explains the general dearth of field BH-low mass star systems
from the computed models.

Dynamical ejection of wide BH-RG systems is further suppressed: owing to the large
envelope size of the RG, an in-cluster
close interaction with such a binary almost inevitably results in
tidal interaction and/or geometrical interception with the giant's envelope,
in turn resulting in a merger, CE, or stable mass transfer with the RG.
Indeed, for the Ba21 clusters, all dynamically paired field BH-RG
systems are stable-mass-transferring. Inspection reveals that
the interactions, that ejected these binaries, are also responsible for
putting them in a mass transferring state. However, it is possible
that an escaped, wide enough ($\porb\sim10^3$ day), eccentric BH-MS binary
evolves into a detached, Gaia BH2-like BH-RG binary.

While no such case is recovered from the Ba21 set, one such field BH-MS binary
has been dynamically produced by a newly computed model cluster set. This set
comprises a grid of model clusters with homogeneous initial conditions
and properties: mass $\mcl=3\times10^4\msun$, size $2$ pc, ``delayed'' remnant
mass model \citep{Fryer12}, metallicities $Z=0.0002$, 0.001, 0.005, 0.01, 0.02,
and overall and massive-star primordial binary fractions (see above) of 10\% and 100\%,
respectively. The distributions of the primordial binaries are similar to
those of Ba21. For each $Z$, model clusters are evolved by
placing them on circular orbits on the equatorial plane of an axisymmetric,
static, Milky Way-like, central mass-disk-halo galactic potential \citep{Allen_1991},
at different galactocentric distances, namely, $\rg=1.0$, 1.5,
2.0, 2.5, 3.0, 3.5, 4.0, 6.0, and 8.5 kpc (hence, the grid contains 45 evolutionary
models in total).
All of these clusters are evolved (using {\tt NBODY7}) until they are
completely dissolved into the external galactic field
(otherwise, at least until all BHs are depleted from them).
A detailed account of this model set (hereafter nicknamed as ``$\rg$-grid'') and its outcomes will
be described in a forthcoming dedicated paper.

As for the BH-star yield, $\rg$-grid has produced one field BH-MS binary with
$\mbh=14.8\msun$, $\mcomp=1.1\msun$, $\porb=1200$ day, $e=0.56$. This binary is
qualitatively similar to the binary demonstrated in Fig.~\ref{Fig:GaiaBH2_evol}
in the latter's BH-MS phase, and would evolve into a Gaia BH2-type detached (non-interacting)
binary. The total simulated initial mass of $\rg$-grid being $\mcltot=1.35\times10^6\msun$,
the unweighted formation rate of field Gaia BH2-type systems from the set is $7.4\times10^{-7}\msun^{-1}$.
These clusters have not produced any field BH-star system that is compatible with
Gaia BH1.

Notably, $\vej$s of the field BH-star binaries with low mass companion are typically lower
than the observed peculiar velocities of the Gaia BH binaries (see Fig.~\ref{fig:Ba21_dists},
lowermost panel). However, the $\vej$ distributions have long tails that extend well beyond
the velocities of both Gaia BH1 and Gaia BH2. In the scenario of dynamical formation
and ejection from moderately massive clusters such as those in Ba21, a generally low
$\vej$ is expected due to the clusters' low velocity dispersion and escape velocity.
Still, high velocity ejections can rarely occur due to chaotic interactions inside
the clusters, invoving compact subsystems. The cluster Gaia BH1 and Gaia BH2 rates quoted
in Table~\ref{tab:rates_clusters} do not include any velocity criterion. 

Since the model clusters produce only one Gaia BH1- and one Gaia BH2-like
binary (see above), we do not repeat the source count analysis in Sec.~\ref{gaiabh_count}
for the cluster channel. We note that
the relative contribution of star clusters to the field Gaia-BH binaries depends
on cluster formation efficiency (hereafter CFE), i.e., the relative fractions of the
star forming mass that go into star clusters and isolated binaries.
An appropriate CFE for a Milky Way-like spiral galaxy is largely
an open question (see, e.g., \citealt{Krumholz_2019} for a review).
The recent study by \citet{DiCarlo23_GaiaBH_form} adopted a CFE of 10\%.
Notably, the number, $\sim900$, of isolated-binary Gaia BH systems, as
obtained in Sec.~\ref{gaiabh_count}, corresponds to the highest formation
efficiency from the isolated-binary channel, which formation efficiency
is similar to that for the cluster channel
(c.f. Tables~\ref{Table:IBE_rates} and \ref{tab:rates_clusters}).
Hence, the 5-year count of Gaia BH-like systems from clusters would 
simply proportionate with the CFE, with respect to the above isolated-binary-channel count.
A detailed study of CFE in the appropriate cluster mass range
is beyond the scope of this work.

\subsection{Discussions: on spin-orbit inclinations of dynamical BH-star binaries}\label{inc_clusters}

As for the spin-orbit alignment, it can be expected that wide, dynamically assembled field
BH-star binaries (resembling Gaia BH1 and BH2; see above)
would have an isotropic spin-orbit misalignment w.r.t. the BH member, since the
exchanging BH is generally uncorrelated to the stellar binary (in a similar
way as for dynamical binary black hole mergers; see, e.g., \citealt{Banerjee_2023}).
In case the pre-exchange (in-cluster) star-star binary is a primordial binary, it may be
spin-orbit aligned to some extent. However, after the exchange, the exchanged BH's spin
would still be uncorrelated to and hence isotropically misaligned w.r.t. the binary's
orbital angular momentum and the stellar companion's spin.
Inside the cluster, close dynamical encounters would impart additional, random, common
spin-orbit tilts to the (BH-star) binary's members \citep{Trani_2021},
so that the isotropic distribution is maintained until the binary's escape from the cluster.

On the other hand, for the very tight field BH-star systems that are formed via in-cluster,
dynamically mediated
tidal interactions (see above;
which are unlike Gaia BH systems), the tidal interaction and circularization may
align the stellar member's spin w.r.t. the orbital angular momentum. However,
the BH member would still remain misaligned, presumably isotropically, at least initially. In the case
of a mass transfer forming an accretion disc around the BH in such a binary,
the system can exhibit low frequency quasi-periodic
oscillations (QPO) related to Lense–Thirring precession of the disc
\citep[][and references therein]{Ingram_2009}.
In other words, stellar clusters may dynamically produce a population of
field stellar mass BH candidates that exhibit potential misalignment signatures
such as low frequency QPOs. Although interesting, a detailed study of this is beyond the scope of this work.

A drawback of the present estimates from clusters is that the criteria for
selecting Gaia BH1 and Gaia BH2 systems could not be matched with
those for the isolated-binary counterpart. This is simply due to the sparsity
of escaped Gaia BH systems (see above), which makes applying well defined
selection criteria nearly impossible. Note that, in this work, we have simply compared
the two formation channels and have not attempted to combine them, unlike
in some recent works \citep{DiCarlo23_GaiaBH_form}.

\begin{table}
\centering
\caption{Formation rates of field BH-star binaries from young massive and open
star clusters. The rates in the first three rows are based on the
evolutionary model star cluster set of Ba21. The Gaia BH2-like systems' rate
in the final row is based on a newly computed cluster model grid (see text).}
\label{tab:rates_clusters}
\begin{tabular}{lcc}
\hline
Case          &      Rate (plain) $[\msun^{-1}]$      & Rate (weighted) $[\msun^{-1}]$\\
\hline
Low mass companion    & $5.8\times10^{-6}$      &     $5.3\times10^{-6}$             \\
Solar mass companion  & $1.4\times10^{-6}$      &     $1.4\times10^{-6}$             \\
Gaia BH1-like         & $2.7\times10^{-7}$      &     $3.5\times10^{-7}$             \\
Gaia BH2-like         & $7.4\times10^{-7}$      &           ---                      \\     
\hline
\end{tabular}
\end{table}

\section{Conclusions}

In this paper we have investigated the two possible channels of the formation of dormant BH binaries Gaia~BH1 and Gaia~BH2: the isolated binary evolution (Sec.~\ref{sec:IBE}) and the dynamical interactions in young massive and open star clusters (Sec.~\ref{GaiaBH_clusters}). We used the population synthesis code $\st$ for IBE calculations (Sec.~\ref{sec:IBE_Method}) and the updated star-by-star N-body integration code {\tt NBODY7} for cluster models (Sec.~\ref{cluster_models}). The two systems from IBE Gaia-BH like binaries population, that match well the parameters of Gaia~BH1 and Gaia~BH2, were evolved till the Hubble time to find the characteristics of the low-mass X-ray binaries that possibly form from such systems (Sec.\ref{ssec:discussion_LMXB}). We have also analysed what are the constrains on magnitude and direction of the natal kick imparted to BH set by the orbital parameters and systemic velocities of Gaia~BH1-like and Gaia~BH2-like binaries (Sec.~\ref{ssec:results_Vsys_Vkick} and Sec.~\ref{ssec:discussion_Vkick}). \\

As seen in Sec.~\ref{res_clusters}, medium mass ($\sim10^4\,\msun$), pc-scale stellar clusters struggle to produce
dynamically paired Gaia BH-type binaries in the field. Nevertheless, the overall population of field BH-star binaries from such clusters does present interesting and diverse demographics. 
In the case when no other constrains ($\porb$, $e$, $M_\mathrm{BH}$) were imposed on the population of binaries except that they should consist of BH and a low-mass star, the formation rate of field BH-star binaries is $1.4-5.8\times10^{-6}\,\msun^{-1}$ depending on the mass of the companion.
This implies that further types of BH-star systems can potentially be discovered in the future. Of course, internally, a cluster continues to assemble BH-star systems (as long as its BH population is not fully depleted), potentially at a much higher rate. The differences between internal and field BH-star population from clusters will be explored in a future work. \\
In contrast, the IBE simulations result in a large Gaia~BH-like binaries population, but this is the effect of pre-choosing possible Gaia~BH-like progenitors as described in Sec.~\ref{ssec:discussion_formation_rates}. Among Gaia~BH-like binaries that are produced through IBE there are systems with the parameters matching closely those inferred from observations of Gaia~BH1 and Gaia~BH2.\\
The results from IBE simulations were obtained using the standard value of parameter $\alpha_\mathrm{ce}$ which describes the fraction of the orbital energy that is used to unbind the envelope in CE phase, that is $\alpha_\mathrm{ce}=1.0$ (see Sec.~\ref{ssec:discussion_lambda}). \\
While we acknowledge the uncertainties that the stellar and binary evolution modelling is subject to, our results indicate that isolated binary evolution should be considered as the scenario equivalent to dynamical interactions in open clusters for Gaia BH1 and Gaia BH2 binaries formation.\\

The formation rate of Gaia~BH-like binaries, as inferred from the present simulations, is sensitive to the criteria adopted to classify a binary as Gaia~BH-like (Sec.~\ref{ssec:results_formation_rates}) and it can differ by three orders of magnitude for the IBE channel (see Table~\ref{Table:IBE_rates}). 
Ensuring that the common criteria are set and the same total stellar mass per simulation is adopted for IBE and dynamical interactions in young clusters, we find that the formation rate of Gaia~BH-like binaries for both channels is equal $2.7\times10^{-7}\,\msun^{-1}$. If we consider the general population of field BH-low mass star ($<3\,\msun$) binaries from young massive and open clusters and corresponding BH-low mass star population from IBE the formation rate of such binaries in both channels is on order of $\sim10^{-6}\,\msun^{-1}$. \\
Our approximate estimate is that $\sim900$ dormant BH binaries, formed through the IBE channel, would be detectable by \textit{Gaia} in the Milky Way thin disc.\\
Considering the IBE channel, the formation of Gaia~BH1- and Gaia~BH2-like binaries depends not only on the natal kick magnitude but also on two angles: the angle between the binary angular momentum and the natal kick vectors ($\gamma$), and the misalignment between the binary angular momentum and the BH-spin vectors ($\theta$). We summarize the results that we inferred from the model Gaia~BH1- and Gaia~BH2-like population as follows:
\begin{itemize}
    \item All Gaia~BH1- and Gaia~BH2-like binaries have systemic velocities $<80$ km/s and the peculiar velocities at birth (which correspond to $v_\mathrm{sys}$ in the model) of the observed systems fit well into the $v_\mathrm{sys}$ distributions of Gaia~BH1- and Gaia~BH2-like binaries.\\ 
    
    \item The natal kick velocity has to be lower than $100$ km/s
    for Gaia~BH1- and Gaia~BH2-like binaries to form. \\
    
    \item The population of Gaia~BH1-like binaries is not vulnerable to the natal kick velocity distribution adopted but a seven fold increase in the number of Gaia~BH2-like binaries in model V2 compared to model V1 favours the two-component Maxwellian distribution with low- and high-velocity components ($\sigma\approx21$ km/s and $\sigma\approx107$ km/s), as suggested by \citet{Zhao2023}.\\
    
    \item The median value of natal kick velocity imparted to the systems, that become Gaia~BH1-like and Gaia~BH2-like binaries, is $\sim39$ km/s for the former and $\sim19$ km/s for the latter (assuming the natal kick distribution from \citet{Zhao2023}).\\
    
    \item $\sim94\%$ of Gaia~BH1-like binaries that form in IBE channel have $\theta<40^{\circ}$ and the median value is $\theta_\mathrm{med}\sim10^{\circ}$.\\
    
    \item For Gaia~BH1-like binary to be formed, the kicks should be directed out of the binary orbital plane: the higher is the natal kick velocity the further from orbital plane should the kick be directed. \\
    
    \item $\sim95\%$ of the binaries that become Gaia~BH2-like experience the natal kick directed within $\pm15^{\circ}$ from the orbital plane.
\end{itemize}

We find that, in addition to the natal kicks, the treatment of the tidal interactions in binaries should not be overlooked in the context of the formation of Gaia~BH-like binaries. The enhancement of the convective damping efficiency relative to the classical prescription \citet{Hut1981}, introduced as the additional factor $F_\mathrm{tid}$ in the equations, increases the number of the systems like Gaia-BH1 in the IBE channel. However, increasing $F_\mathrm{tid}$ above the certain value, which was originally calibrated to match the observations \citep[$F_\mathrm{tid}=50$,][]{Belczynski08}, does not lead to further significant increase in the Gaia~BH1-like binaries number (see Sec. \ref{sec:discussionIBE_tidal}). The results show that the prescriptions for tidal interactions that are used in the codes play a decisive role in the formation of Gaia~BH1-like binaries via the IBE channel and that more detailed investigation on the subject is needed.
\\
Tracking the future evolution of Gaia~BH-like binaries shows that those systems may be the progenitors of LMXBs. Those LMXBs would be transient systems with the recurrence times of order of tens to over hundred years according to our calculations.

\section*{Acknowledgements}
The co-author, Prof. Krzysztof Belczynski, passed away on January 13th 2024. He contributed to this paper with his ideas and discussions.\\
I.K. would like to thank Prof. Jean-Pierre Lasota for his comments that helped to improve the manuscript and to Ataru Tanikawa and Aleksey Generozov for helpful discussions.\\
We thank the referee for their constructive report that helped to improve the manuscript.\\
I.K. and K.B are supported by the Polish National Science Center (NCN) grant Maestro (2018/30/A/ST9/00050).
S.B. acknowledges funding for this work by the Deutsche Forschungsgemeinschaft (DFG, German Research Foundation)
through the project ``The dynamics of stellar-mass black holes in dense stellar systems and their
role in gravitational wave generation'' (project number 405620641; PI: S. Banerjee).
Special thanks go to tens of thousands of
citizen-science project "Universe@home" (universeathome.pl) enthusiasts that help to develop the
$\st$ population synthesis code used in this study. 

\section*{Data Availability}
The data presented in this work can be made available based on the
request to the corresponding author.

\bibliographystyle{mnras}
\bibliography{GaiaBH}

\appendix

\section{}\label{Sec:App}

\begin{table}]\label{table:app_trecc}
    \centering
    \caption{The parameters used to calculate the outburst recurrence times at the onset and finish of LMXB phase.}  
  	\label{Table:Trecc_param}
	\begin{tabular}{c|c|c} 
         \hline
                                                 & onset     & finish  \\
          \hline                       
		$M_\mathrm{BH}$ [\(\msun\)]          & 9.7    & 10.2 \\
		$\dot{M}_\mathrm{tr}$ [g/s]          & $4.45\times10^{18}$ & $6.07\times10^{18}$  \\
            $R_\mathrm{d}$ [cm]                  & $9.61\times10^{12}$ & $2.73\times10^{13}$ \\  
            \hline
	\end{tabular}
\end{table}

\section{Gaia~BH1-like binaries as LISA sources}\label{ssec:diss_lisa}

We checked if Gaia~BH1-like binaries become LISA sources at some point of their late evolution using the parameters of a binary described in Sec.~\ref{ssec:results_evolution}. We calculate the gravitational waves (GW) strain amplitude $h$ taking into account that Gaia~BH1 will become BH-WD binary according to our results. We start from the formula for the gravitational wave strain derived by \citet{Nelemans01}:

\begin{equation}
    h=1.0\times10^{-21}\frac{\sqrt{g(n,e)}}{n}\left( \frac{M_\mathrm{Ch}}{\mathrm{M}_\odot}\right)^{5/3}\left(\frac{P_\mathrm{orb}}{1\,\mathrm{hr}}\right)^{-2/3}\left(\frac{d}{1\,\mathrm{kpc}}\right)^{-1}, 
\end{equation}
where $g(n,e)$ is the Fourier decomposition of GW signal to $n$ harmonics, given by the combination of Bessel functions of first kind \citep[see][Eq.10]{Barack04}, $M_\mathrm{Ch}$ is a chirp mass in solar masses defined as $M_\mathrm{Ch}=(M_\mathrm{BH}M_2)^{3/5}(M_\mathrm{BH}+M_2)^{-1/5}$, $P_\mathrm{orb}$ is the orbital period in hours and $d$ is the distance to the binary in kilo-parsecs.\\
For circular binaries the factor $g(n,e)$ is non-zero only when $n=2$ ($g(2,0)=1.0$), therefore, the frequency of GW emitted by circular binaries is always twice their orbital frequency. 
We calculate the GW strain $h(n,e)$ for $n=2$, $e=0.0$, $M_\mathrm{BH}=10.2\,$ \(\msun\), $M_2=0.5$ \(\msun\), $P_\mathrm{orb}=979.6$ days and $d=0.48$ kpc, that is when Gaia~BH1 becomes BH-WD binary:

\begin{equation}
    h(2,0)=2.03\times10^{-23} \,\,\,\,\,\,\, f_\mathrm{GW}=4.99\times10^{-7}  \mathrm{Hz}
\end{equation}

Comparing the result with the expected LISA sensitivity \citep{Lisa_Curves2019} it is clear that Gaia~BH1 will not contribute to LISA sources. \\
The Gaia~BH2 parameters, when it becomes BH-WD binary, are comparable to Gaia~BH1 except for the orbital separation which is $\sim2.5$ times wider. It is apparent that GW emitted from Gaia~BH2 will be undetectable by LISA likewise, therefore we skip the calculations of GW strain for Gaia~BH2.

\bsp	
\label{lastpage}
\end{document}